\documentclass{aa}  

\usepackage{pdflscape}
\usepackage{graphicx}
\usepackage{float}
\usepackage{txfonts}

\usepackage{bm}
\usepackage{longtable}
\usepackage{booktabs}
\usepackage{lscape}
\usepackage{array}
\usepackage{hyperref}

\newcolumntype{C}[1]{>{\centering\arraybackslash}p{#1}}

\usepackage{natbib}
\bibpunct{(}{)}{;}{a}{}{,}             

\usepackage{color}

\begin{document}

	\title{Cosmographic constraints from late-time probes including fast radio bursts}
	
	
	\author{Lázaro L. Sales
		\inst{1,2}\fnmsep\thanks{Corresponding author: lazarolima@uern.br}
		\and
		Klecio E. L. de Farias\inst{2}
		\and
		Amilcar R. Queiroz\inst{2}
		\and
		Joao R. L. Santos\inst{2,3,9}
		\and
		Rafael A. Batista\inst{2}
		\and
		Ana R. M. Oliveira\inst{4}
		\and
		Lucas F. Santana\inst{2}
		\and
		Carlos A. Wuensche\inst{5}
		\and
		Thyrso Villela\inst{5,6,7}
		\and
		Jordany Vieira\inst{8}
	}
	
	\institute{Departamento de F\'{\i}sica, Universidade do Estado do Rio Grande do Norte, 59610-210, Mossor\'o, RN, Brazil
		\and
		Departamento de F\'{\i}sica, Universidade Federal de Campina Grande, Caixa Postal 10071, 58429-900, Campina Grande, Para\'{\i}ba, Brazil.
		\and
		Unidade Acad\^emica de Matem\'atica, Universidade Federal de Campina Grande, 58429-970,  Campina Grande, Para\'{\i}ba, Brazil.
		\and
		Departamento de F\' isica, Universidade Federal da Para\' iba,  Caixa Postal 5008, Jo\~ ao Pessoa, Para\' iba, Brazil.
		\and
		Instituto Nacional de Pesquisas Espaciais (INPE), Divisão de Astrofísica, Av. dos Astronautas 1758, Jardim da Granja, São José dos Campos, SP, Brazil.
		\and
		Centro de Gestão e Estudos Estratégicos (CGEE), SCS, Qd. 9, Lote C, Torre C S/N, Salas 401 a 405, 70308-200, Brasília, DF, Brazil.
		\and
		Universidade de Brasília (UnB), Instituto de Física, Campus Universitário Darcy Ribeiro, 70910-900, Brasília, DF, Brazil.
		\and
		Instituto de Física, Universidade de São Paulo, C.P. 66.318, CEP 05315-970, São Paulo, Brazil.
		\and
		Institute for Theoretical Physics, Leibniz University Hannover, Appelstraße 2, 30167 Hannover, Germany.
	}
	
	
	
	\abstract
	{Late-time astrophysical probes offer a powerful way to investigate the expansion history of the Universe. Among these probes, fast radio bursts (FRBs) are millisecond-duration astrophysical transients whose extragalactic origin makes them a promising addition to studies of large-scale structure. Their dispersion measures encode information about the intervening cosmic plasma, which can be linked to cosmological distances.}
	{In this study, we use late-time probes, such as well-localized FRBs, baryon acoustic oscillations (BAO), supernovae (SNe), and cosmic chronometers (CC) to constrain cosmological parameters through a model-independent cosmographic approach.}
	{By integrating FRB data with BAO from DESI DR2, SNe, and CC, we derive constraints on the Hubble constant ($H_0$), the deceleration parameter ($q_0$), and the jerk parameter ($j_0$), using Markov Chain Monte Carlo (MCMC) analysis for parameter estimation.}
	{The cosmographic approach with FRBs alone provides 
		$H_0 = 66.35^{+4.13}_{-5.04} \, \text{km} \, \text{s}^{-1} \, \text{Mpc}^{-1}$, 
		$q_0 = -0.33^{+0.21}_{-0.15}$, 
		and 
		$j_0 = 0.83^{+0.57}_{-0.67}$, 
		corresponding to a precision of $\sim 6\%$ for the Hubble constant and showing consistency with the $\Lambda$CDM expectation. 
		The DESI+CMB dataset yields 
		$H_0 = 65.59^{+1.25}_{-1.24} \, \text{km} \, \text{s}^{-1} \, \text{Mpc}^{-1}$, 
		$q_0 = -0.29^{+0.07}_{-0.08}$, 
		and 
		$j_0 = 0.58^{+0.03}_{-0.04}$, 
		providing a $\sim 2\%$ precision on $H_0$ and may suggest a possible tension in the late-time kinematic sector relative to the $\Lambda$CDM expectation when BAO measurements are calibrated with a Planck-inferred sound horizon. Combining the FRB, SNe, DESI+CMB, and CC datasets further tightens the constraints to 
		$H_0 = 68.03^{+0.53}_{-0.52} \, \text{km} \, \text{s}^{-1} \, \text{Mpc}^{-1}$, 
		$q_0 = -0.41 \pm 0.02$, 
		and 
		$j_0 = 0.55 \pm 0.02$, 
		with the jerk parameter remaining lower than $j_0 = 1$ at the $1\sigma$ confidence level.}
	{These findings hint at a possible late-time kinematic tension, as indicated by the inferred value of the jerk parameter, which is primarily driven by the DESI+CMB dataset under standard early-Universe assumptions for the sound horizon. At the current level of observational precision, FRBs play a complementary role in the cosmographic analysis, with their impact expected to increase as larger and more precise samples of well-localized events become available.}

	
	\maketitle
	%
	
	\section{Introduction}
	
	On July 24, 2001, the Parkes radio telescope detected a very bright and unidentified radio signal in the $1.4\,\mathrm{GHz}$ survey of the Magellanic Clouds. The burst had a significant flux of $30\pm10\,\mathrm{Jy}$,  with $<5\,\mathrm{ms}$ duration and was reported by Lorimer and collaborators in 2007 \citep{Lorimer:2007qn}. These astronomical events are known as Fast Radio Bursts (FRBs), and their population has been increasing, reaching about 1,000 events at the time of writing \footnote{Data obtained from the Transient Name Server: \url{https://www.wis-tns.org/}}. One of the key properties of FRBs is their dispersion measure (DM), which refers to the time delay of the FRB signal across different radio frequency channels. This delay occurs because of interactions with free electrons along the signal's path, causing lower-frequency wave fronts to arrive later than higher-frequency ones. By analyzing this time delay, we can estimate the distance of the FRB and assess the matter content along the line of sight. Because FRBs exhibit an extremely high dispersion that surpasses the maximum contribution from the Milky Way, they are classified as extragalactic events. 
	
	The classification of FRBs can be divided into repeaters (periodic or not) and non-repeaters, the latter corresponding to a single detected pulse. The majority of FRBs detected across different surveys are non-repeating. Despite numerous efforts to identify their origin, the sources of these phenomena remain unclear. However, repeating signals may be attributed to neutron stars with strong magnetic fields, such as pulsars \citep{Cordes:2015fua} and magnetars \citep{Popov:2007uv,Pen:2015ema}. However, since most FRBs are extragalactic in origin, identifying the host galaxy of a detected FRB may help constrain cosmological models \citep[e.g.,][]{kalita2024fast} and estimate the fraction of baryonic mass in the intergalactic medium (IGM) \citep{Munoz:2018mll,Walters:2019cie,Qiang:2020vta}.
	
	FRBs with well-defined distances offer a promising pathway for studying open problems in astrophysics and cosmology, such as the Hubble tension \cite[see][]{james2022measurement,gao2024measuring,kalita2024fast,connor2025gas}. The investigation of Hubble tension can be addressed through various methods and theories \citep{di2021realm}. 
	
	In this work, motivated by the Hubble tension and recent indications of deviations from the standard cosmological model reported by the DESI collaboration, we explore a model-independent approach based on a kinematic model, also known as cosmography \citep{Gao:2014iva,gao2024measurement}. The only constraints applied to this method are the FLRW metric and the cosmological principle. Therefore, it is largely used to derive model-independent cosmological parameters \citep{Visser:2003vq,Gao:2014iva,Aviles:2016wel,gao2024measurement}. In this method, only the evolution of the expansion from direct measurements of the kinematic parameters is considered, so we do not need to worry about the composition of the Universe. Moreover, this approach allows us to derive several cosmological parameters, such as the Hubble parameter $H_0$, deceleration parameter $q_0$, and jerk parameter $j_0$, among other contributions stemming from the high-order derivatives of the scale factor. In this context, our MCMC routine is used to estimate the cosmographic parameters $H_0$, $q_0$, and $j_0$ with late-time observational probes. In this approach, the observables for FRB, BAO, SNe, and CC are expressed in terms of the cosmographic parameters. We then compare the constraints obtained from the FRB data with those derived from SNe, CC, and the latest results from the DESI collaboration.
	
	The datasets applied to constrain the cosmographic parameters in our analysis are: the Pantheon+SH0ES for 1701 SNe Ia light curves in the redshift range $0.001<z<2.26$ \citep{Brout:2022vxf,Scolnic:2021amr,riess2022comprehensive}; the CC with 35 observed data in the redshift range $0.07<z<1.965$ \citep{Hu:2024big}; DESI DR2 Results II, provided by BAO measurements of over 14 million galaxies and quasars in the redshift range $0.1<z<4.2$ \citep{DESI:2025zpo,DESI:2025zgx}; and a sample of 116 localized FRBs \citep{Acharya:2025ubt,Wang:2025ugc}. We also used the Planck 2018 results \citep{aghanim2020planck} for some of our priors in the MCMC and for comparing the values of $H_0$ constrained with our methods.  
	
	For easy reading, this manuscript is divided into the following sections. In Sec. \ref{sec2}, we present a brief overview of the FRB theory. Theoretical approaches are discussed in Sec. \ref{sec3}, where the cosmography is explained in detail. In Sec. \ref{datasets}, the datasets are presented to ensure compatibility with each other and the applied method. Statistical and computational procedures are outlined in Sec. \ref{methods}. The results are reported in Sec. \ref{sec7}, where the cosmographic parameters $q_0$ and $j_0$, as well as the Hubble constant $H_0$, are computed using several datasets, including the FRB sample and the new results of DESI. Lastly, in Sec. \ref{conclusion}, we present our final remarks, highlighting the main results obtained in this work.
	
	
	\section{Fast radio bursts}
	\label{sec2}
	
	Fast radio bursts are intense, millisecond-duration radio transients of extragalactic origin, first discovered unexpectedly in 2007. They are characterized by their high flux densities (typically $0.5$--$100$~Jy), very short durations (a few milliseconds or less), and large DM that often exceed the expected Galactic contribution along the line of sight. These features suggest that FRBs originate at cosmological distances, with inferred luminosities reaching up to $\sim 10^{43}$ erg s$^{-1}$. Their emission is coherent and spans a broad frequency range, from $\sim 400$~MHz up to $\sim 8$~GHz, although the precise spectral behavior often deviates from a simple power law and can be highly variable across bursts. The observed FRB population includes both one-off events, possibly of a cataclysmic nature, and repeating sources, implying diverse progenitor classes or emission mechanisms. In particular, the repeating FRB~121102 shows extreme rotation measures and complex spectral and temporal substructure, suggesting a dense and magnetized local environment \citep{michilli2018extreme}.
	
	The main observational properties of FRBs include significant frequency-dependent dispersion, temporal scattering, and in some cases, scintillation and strong Faraday rotation, all of which offer important diagnostics of the intervening medium. Pulse widths range from sub-milliseconds to tens of milliseconds and often exhibit asymmetric temporal profiles due to multi-path propagation effects. The spectral indices are poorly constrained due to limited bandwidth and uncertain source localization within the telescope beam. Some FRBs are highly polarized, with rotation measures exceeding $10^5$~rad~m$^{-2}$, indicating extreme magneto-ionic environments. The diversity of pulse shapes, spectral variability, and propagation effects complicates the classification and interpretation of FRBs but also provides an opportunity to use them as probes of the interstellar and intergalactic media. Their high brightness temperature ($T_B \sim 10^{35} - 10^{37}$~K) and short durations require coherent emission processes, distinguishing them from incoherent synchrotron sources. As the FRB detection rate increases with new wide-field radio facilities, statistical analyses of their population properties will become a powerful cosmological and astrophysical tool. For a review of the properties of FRBs, interested readers can refer to Refs. \citep{Petroff:2019tty,cordes2019fast}
	
	The DM is crucial for constraining cosmological models using FRB data. It accounts for contributions from various components, such as the host galaxy (where the FRB is supposedly located), the intergalactic medium and the interstellar medium of our galaxy. The total DM contribution, from the source to the observer, is calculated using the following expression \citep{Petroff:2014taa}:
	\begin{equation}
		{\rm DM}_{\mathrm{obs}}(z)= {\rm DM}_{\mathrm{MW}}+{\rm DM}_{\mathrm{IGM}}+\frac{{\rm DM}_{\textrm{host}}}{1+z}\;,
		\label{e1}
	\end{equation}
	where the contribution of Milky Way reads as ${\rm DM}_{\rm MW} = {\rm DM}_{\rm MW,ISM}+{\rm DM}_{\rm MW,halo}$ with DM$_{\textrm{MW,ISM}}$ being the galactic interstellar medium $(\sim 10^0-10^3\, {\rm pc\ cm^{-3}})$ and DM$_{\textrm{MW,halo}}$ the galactic halo. DM$_{\textrm{IGM}}$ denotes the intergalactic medium $(\sim 10^2-10^3\, {\rm pc\ cm^{-3}})$ and DM$_{\textrm{host}}$ the host galaxy interstellar medium $(\sim 10^0-10^3\, {\rm pc\ cm^{-3}})$. The Milky Way halo contribution DM$_{\textrm{MW,halo}}$ is not well defined, and the expectation values lie between $50-100\, {\rm pc\ cm^{-3}}$. The factor $(1+z)$ in \eqref{e1} is due to cosmic dilation.

	When an FRB pulse approaches Earth, it interacts with all the components of Eq. \ref{e1}, resulting in dispersion and a time delay between the arrival of different frequencies in the observed signal. This delay is characterized by \citep{Petroff:2019tty}
	\begin{equation}
		\Delta t \propto \left(\frac{1}{\nu_{\rm lo}^2} - \frac{1}{\nu_{\rm hi}^2}\right){\rm DM}\;,
	\end{equation}
	where $\nu_{\rm lo}$ and $\nu_{\rm hi}$ stand for the lower and higher frequencies of the emitted signal. The DM parameter relates to the column density of free electrons ($n_e$) along the line of sight ($l$) to the FRB \cite[see, e.g.,][]{Petroff:2019tty}), so that 
	\begin{equation}
		{\rm DM} = \int \frac{n_e(z)dl}{1+z}\;.
	\end{equation}
	
	To use FRB data as a cosmological probe, we must account for the observed extragalactic DM, which is determined by removing the Milky Way contribution from the observed DM, i.e., 
	\begin{align} 
		{\rm DM}_{\rm ext}(z) &\equiv {\rm DM}_{\mathrm{obs}}(z)-{\rm DM}_{\mathrm{MW}}\nonumber\\
		&={\rm DM}_{\mathrm{IGM}}+\frac{{\rm DM}_{\textrm{host}}}{1+z}\;.
		\label{DM_ext_th}
	\end{align}
	The DM accumulated along a cosmological line of sight reflects the column density of free electrons from diffuse baryons in the IGM, whose distribution in space is highly inhomogeneous due to the cosmic large-scale structure. In the analysis by \citet{Macquart:2020lln} the probability distribution for ${\rm DM}_{\mathrm{IGM}}$, predicted from semi-analytic models and simulations, is well approximated by an analytic form that is nearly Gaussian in the bulk but with an extended tail towards high \( \mathrm{DM} \) values, capturing the rare but significant contributions from dense intervening structures (e.g., voids, galaxy halos or filaments). This supports the use of a quasi-Gaussian distribution -- roughly Gaussian around the mean DM but allowing for an asymmetric, long tail -- as a physically motivated model for the ${\rm DM}_{\mathrm{IGM}}$. In this regard, we adopt the following form for the PDF of $\rm DM_{\rm IGM}$ \citep{Macquart:2020lln}:
	\begin{equation}
		P_{\text {IGM}}(\Delta)=A \Delta^{-\beta} \exp \left[-\frac{\left(\Delta^{-\alpha}-C_0\right)^2}{2 \alpha^2 \sigma_{\text {IGM }}^2}\right], \quad \Delta>0
		\label{prop1}
	\end{equation}
	where
	\begin{equation}
		\Delta \equiv \frac{\mathrm{DM}_{\mathrm{IGM}}}{\left\langle\mathrm{DM}_{\mathrm{IGM}}\right\rangle}=\frac{\mathrm{DM}_{\mathrm{ext}}-(1+z)^{-1}\mathrm{DM}_{\mathrm{host}}}{\left\langle\mathrm{DM}_{\mathrm{IGM}}\right\rangle}~.
		\label{prop2}
	\end{equation}
	Here, $A$ is a normalization constant and $C_0$ is determined by requiring $\left\langle\Delta\right\rangle=1$. Following \citet{Macquart:2020lln}, we use $\alpha=\beta=3$ and $\sigma_{\rm IGM}=Fz^{-0.5}$ with $F=0.32$.
	
	The average DM for the DM$_{\rm IGM}$ in terms of the redshift is described by the Macquart relation \citep{Macquart:2020lln}, whose form is
	\begin{equation} 
		\langle {\rm DM}_{\rm IGM}(z) \rangle = A \Omega_b H_0^2 \int_0^z \frac{\left(1+z^{\prime}\right) x_e\left(z^{\prime}\right)}{H\left(z^{\prime}\right)} d z^{\prime}~,
		\label{e2}
	\end{equation}
	with $A=3cf_{\rm IGM}/(8 \pi G m_p)$, where $G$ is Newton's gravitational constant, $f_{\rm IGM}=0.83$ is the baryon mass fraction present in the IGM \citep{shull2012baryon}, and $H(z)$ is the Hubble parameter, carrying all the cosmological information. In order to use FRB data to constrain cosmological parameters, let us adopt the cosmography approach for $H(z)$ presented in Eq. \eqref{e2}. Such an approach aims to alleviate the dependence on a specific cosmological model (see section \ref{sec3} for more details). The function $x_e(z)$ in Eq. \eqref{e2} denotes the free electron fraction (or degree of plasma ionization), and it is given by 
	\begin{equation}
		x_{\rm e}(z)=Y_{\rm H} x_{\rm e, H}(z)+\frac{1}{2}Y_{\rm He} x_{\rm e, He}(z)\;, 
	\end{equation}
	with $Y_{\mathrm{H}}=3/4$ and $Y_{\mathrm{He}}=1/4$ being the mass fractions of hydrogen and helium, respectively. In addition, $x_{\mathrm{e}, \mathrm{H}}(z)$ and $x_{\mathrm{e}, \mathrm{He}}(z)$ denote the ionization fractions of hydrogen and helium, respectively. In our analysis, we shall consider $x_{\rm e, H}(z)=x_{\rm e, He}(z)=1$ since hydrogen and helium are fully ionized at $z<3$ \citep{Meiksin:2007rz,becker2011detection},  which gives $x_{\rm e}=7/8$. 
	
	Conversely, the DM contribution from the host galaxy, denoted as $\mathrm{DM}_{\mathrm{host}}$, arises from the integrated column density of free electrons within the galaxy. This contribution is influenced by several factors, including the galaxy's morphology, inclination, star formation rate, and the FRB's location within the host. These factors introduce significant variability and skewness in the $\mathrm{DM}_{\mathrm{host}}$ distribution. Cosmological hydrodynamical simulations, such as those from IllustrisTNG, have demonstrated that the distribution of $\mathrm{DM}_{\mathrm{host}}$ can be effectively modeled by a log-normal function across different redshifts and galaxy types \citep{zhang2020dispersion}. This approach captures the asymmetric nature of the distribution and provides a practical framework for interpreting FRB observations and constraining host galaxy properties. Based on the above discussion, in this work, we will adopt a log-normal distribution for the $\mathrm{DM}_{\mathrm{host}}$,
	\begin{align} \label{p_host}
		P_{\rm host}(\mathrm{DM}_{\mathrm{host}}|z) =& \frac{1}{\sqrt{2\pi} \, \sigma_{\mathrm{host}} \, \mathrm{DM}_{\mathrm{host}}}\nonumber\\ 
		&\times\exp\left[-\frac{1}{2} \left( 
		\frac{\ln(\mathrm{DM}_{\mathrm{host}}) - \mu}{\sigma_{\mathrm{host}}}
		\right)^2 \right]~,
	\end{align}
	where $\mu = \langle {\rm DM}_{\rm host} \rangle$ and $\sigma_{\rm host}$ are both the log-normal distribution parameters. In our analysis, such parameters will be treated as free parameters. 
	
	\section{Cosmography} \label{sec3}
	
	An alternative approach to investigate the late-time evolution of the Universe is the kinematic model, also called cosmography \citep{Gao:2014iva}. In this method, only the evolution of the expansion from direct measurements of the kinematic parameters is considered, meaning that we do not need to worry about the composition of the Universe. This approach has the advantage of being model-independent since the parameters are obtained directly from observational data. To obtain the cosmographic parameters, we start with the Hubble parameter $H(t)$, expressed as
	\begin{equation}
		H(t)=\frac{\dot{a}}{a}\,,
		\label{c1}
	\end{equation}
	where $a(t)$ is the scale factor that describes the expansion of the Universe as a function of time. By performing a Taylor expansion around the scale factor, we get 
	\begin{align}
		a(t)=&a_0\left\{1+  H_0\left(t-t_0\right)-\frac{1}{2} q_0 H_0^2\left(t-t_0\right)^2+\frac{1}{3!} j_0 H_0^3\left(t-t_0\right)^3\right.\nonumber\\
		&\left.+\mathcal{O}((t - t_0)^4)\right\}\,,
		\label{c2}
	\end{align}
	where $q_0$ and $j_0$ are the deceleration and the jerk cosmographic parameters. These parameters are defined as  
	\begin{equation}
		q_0 = -\frac{\ddot{a}(t_0)}{a(t_0) H_0^2}~~~~{\rm and}~~~~j_0 = \frac{\dddot{a}(t_0)}{a(t_0) H_0^3} \,.
		\label{c3}
	\end{equation}
	Here, $q_0$ stands for the behavior of cosmic expansion; a positive value means a slowed-down expansion and a negative one indicates an accelerated expansion. While $q_0$ describes whether the Universe is accelerating or decelerating, $j_0$ captures the rate of change of the acceleration in the Universe's expansion. In the realm of $\Lambda$CDM model, where the accelerated expansion is driven by the cosmological constant, $j_0$ is equal to 1. On the other hand, deviations of $j_0=1$ suggest non-standard cosmologies, potentially involving dynamic dark energy \citep{Enkhili:2024dil}, modified gravity \citep{Myrzakulov:2023ohc}, or other phenomena \citep{Demianski:2016dsa}. 
	
	The Hubble parameter can be written in terms of the above parameters as follows:
	\begin{equation} \label{H_cosmo}
		H(z) = H_0 \left[ 1 + (1 + q_0) z + \frac{1}{2} \left(j_0 - q_0^2\right) z^2 + \mathcal{O}(z^3) \right]\,.
	\end{equation}
	The dependence on the luminosity distance $d_L(z)$ of the Hubble parameter is reflected in a description of this quantity in terms of $q_0$ and $j_0$ \citep[see details in][]{Visser:2003vq}, whose form is
	\begin{align} \label{dL_cosmo}
		d_{\rm L}(z) =& \frac{c}{H_0} \left[ z + \frac{1 - q_0}{2} z^2 \right.\nonumber\\
		&\left.+ \frac{1}{6} \left(3 q_0^2 + j_0 + q_0 - 1\right) z^3 + \mathcal{O}(z^4) \right]\,.
	\end{align}

	To address FRB physics in the context of cosmography, we need to know how the DM of the intergalactic medium is affected by the $q_0$ and $j_0$ cosmographic parameters.  The ${\rm DM}_{\rm IGM}$ from \eqref{e2} can be approximated by
	\begin{align}
		\langle \mathrm{DM}_{\mathrm{IGM}}(z)\rangle =& 10^{4}A x_{\rm e} \Omega_{\rm b} h^2\frac{1}{H_0} \left[ z - \frac{q_0 z^2}{2}\right.\nonumber\\
		&\left.+ \frac{1}{6} \left( 4 + 6 q_0 + 3 q_0^2 - j_0 \right) z^3 + \mathcal{O}(z^4) \right] ~.
		\label{c15}
	\end{align}
	
	Furthermore, to include BAO in our analysis within the framework of cosmography, the relevant equations are
	\begin{align} \label{D_M}
		D_{\rm M}(z) =& \frac{c}{H_0}\left\{z - \frac{(1+q_0)z^2}{2}\right.\nonumber\\
		&\left.+ \frac{1}{3}\left[(1+q_0)^2 - \frac{1}{2}(j_0 - q_0^2)\right]z^3 + \mathcal{O}(z^4)\right\},
	\end{align}
	which is the  transverse comoving distance for a flat Universe, and the Hubble distance 
	\begin{align} \label{D_H}
		D_{\rm H}(z) =& \frac{c}{H_0}\left\{1 - (1+q_0)z\right.\nonumber\\
		&\left.+ \left[(1+q_0)^2 - \frac{1}{2}(j_0 - q_0^2)\right]z^2 + \mathcal{O}(z^3)\right\}~.
	\end{align}
	Since inferred distances are relative to the sound horizon ($r_{\rm d}$), the directly constrained quantities are the ratios $D_{\rm M}/r_{\rm d}$, $D_{\rm H}/r_{\rm d}$ and $D_{\rm V}/r_{\rm d}$, with $D_{\rm V}=[zD_{\rm M}^2(z)D_{\rm H}(z)]^{1/3}$ being the angle-averaged distance.  
	
	Although additional cosmographic parameters, such as $s_0$ and $l_0$, are included in $\mathcal{O}(z^n)$, we will focus solely on $H_0$, $q_0$, and $j_0$. The cosmographic approach exhibits linear behavior at low redshifts; however, at higher redshifts, parameters derived from higher-order terms of the expansion become increasingly significant. Nevertheless, due to the mathematical limitations of the Taylor expansion in terms of redshift $z$, which has a convergence radius of at most $|z| = 1$ \citep{cattoen2007hubble}, we restrict our analysis to data with $z < 1$ by applying a redshift mask to all observational samples here considered. This ensures that the series expansion remains valid within its domain of convergence. Furthermore, we truncate the expansion at second order -- i.e., including terms up to the jerk parameter $j_0$ -- to balance precision and reliability. Adding higher-order terms may improve the accuracy of the approximation, but such a procedure introduces substantial degeneracy among the parameters, thereby undermining the stability and interpretability of the inference. We aim to preserve physical coherence and statistical robustness in our cosmographic analysis by limiting the expansion to the second order.
	
	\section{Datasets} \label{datasets}
	
	To provide consistent constraints on the cosmological parameters studied, three additional observational datasets are used to enhance the results obtained from the FRB dataset. The details on these datasets are presented below:
	
	\begin{itemize}
		
		\item Among the new set of experiments mapping the large scale of the Universe and characterizing the dark energy behavior in different redshifts is the Dark Energy Spectroscopic Instrument (DESI). DESI started its operation in 2021 and recently released new data results \citep{DESI:2025zgx,DESI:2025zpo}, based on BAO measurements carried out over three years of observations. The DESI collaboration aims to constrain cosmological models, particularly dark energy and neutrino masses, using over 14 million galaxies and quasars, covering the redshift range $0.1<z<4.2$. This dataset includes Bright Galaxy Samples (BGS), Luminous Red Galaxies (LRGs), Emission Line Galaxies (ELGs), quasars (QSOs), and Ly$\alpha$ forest measurements. In this work, we use the DESI DR2 Results II, which comprise 13 correlated measurements\footnote{\url{https://github.com/CobayaSampler/bao_data/tree/master/desi_bao_dr2}}
		of $D_{\rm M}/r_{\rm d}$, $D_{\rm H}/r_{\rm d}$, and $D_{\rm V}/r_{\rm d}$.
		
		\item We use the Pantheon+SH0ES dataset, which contains 1550 unique astronomical sources contributing to a total of 1701 type Ia supernova light curves. This dataset presents a wide redshift range from $0.001<z<2.26$, enabling detailed cosmological analysis of the expansion history. This compilation combines improved calibration, photometry, and light-curve fitting procedures from multiple surveys \citep{Brout:2022vxf,Scolnic:2021amr,riess2022comprehensive}.
		
		\item We also include data from the cosmic chronometers method, consisting of 35 observational points in the redshift range $0.07<z<1.965$. These measurements are based on the differential age evolution of massive galaxies and provide a model-independent estimate of the Hubble parameter $H(z)$. Here we adopt the data points from Ref. \citet{Hu:2024big}.
		
		\item The FRB data consists of 116 localized FRBs. This sample was derived from Refs. \citet{Acharya:2025ubt,Wang:2025ugc}, which selected the intersection of both datasets, resulting in the current sample. The authors compiled a table from several radio telescopes. In Appendix \ref{append_a}, Table \ref{tab1} presents the FRB sample used in the current analysis. Furthermore, the value of DM$_{\rm MW}$ for each burst is obtained using the NE2001 model, as in \citet{Cordes:2002wz}, and we do not distinguish between repeating and non-repeating FRBs.
	\end{itemize}
	
	Since the convergence of the cosmographic expansion requires $z<1$, when constraining cosmographic parameters, we apply a mask to filter all samples considered in this work to $z<1$. As a result, the FRB sample will be reduced to 114 data points, SNe to 1563 points, BAO to 7 points, and CC to 27 points.

	\section{Methods} \label{methods}
	
	\subsection{Likelihoods and Bayesian inference}
	
	In this subsection, we present the statistical approaches adopted in our analysis. We aim to constrain the cosmological parameters using cosmography. By employing this method, we strive to alleviate the dependence on a specific cosmological model, avoiding issues related to degeneracy between the model's parameters. This study explores how cosmography can constrain cosmological parameters using FRB, SNe, BAO, and CC data, and their combinations. 
	
	To construct the likelihood function of FRBs, we use Eq. \eqref{DM_ext_th} and the definition given by \eqref{prop2} to find
	\begin{align}
		P_{\mathrm{IGM}}\left(\mathrm{DM}_{\mathrm{ext}}-(1+z)^{-1}\mathrm{DM}_{\mathrm{host}}\right)  =P_{\mathrm{IGM}}\left(\Delta \times\left\langle\mathrm{DM}_{\mathrm{IGM}}\right\rangle\right)&\nonumber \\
		=\frac{1}{\left\langle\mathrm{DM}_{\mathrm{IGM}}\right\rangle} P_{\text {IGM}}\left(\frac{\mathrm{DM}_{\mathrm{ext}}-(1+z)^{-1}\mathrm{DM}_{\text {host }}}{\left\langle\mathrm{DM}_{\mathrm{IGM}}\right\rangle}\right)&\,,
		\label{prop3}
	\end{align}
	where the product of two random variables $z=xy$ for a PDF is \citep{rohatgi2015introduction}
	\begin{equation}
		P(z)=\int_{-\infty}^{\infty} \frac{1}{|x|} \times P_x(x) \times P_y\left(\frac{z}{x}\right) d x
		\label{prop4}
	\end{equation}
	Consequently, 
	\begin{align}
		P\left(\mathrm{DM}_{\text {ext }}\right)=&\frac{1}{\left\langle\mathrm{DM}_{\mathrm{IGM}}\right\rangle}\int_0^{\mathrm{DM}_{\text {ext }}(1+z)} P_{\text {host }}\left(\mathrm{DM}_{\text {host }}\right) \nonumber\\
		&\times P_{\rm IGM}\left(\frac{\mathrm{DM}_{\text {ext }}-(1+z)^{-1}\mathrm{DM}_{\text {host }}}{\left\langle\mathrm{DM}_{\mathrm{IGM}}\right\rangle}\right) d \mathrm{DM}_{\text {host }} .
		\label{prop5}
	\end{align}
	This corrected expression was obtained by  \citet{Zhang:2025wif} where they revisited the expression from \citet{Macquart:2020lln} by introducing a multiplicative factor $1/\left\langle\mathrm{DM}_{\mathrm{IGM}}\right\rangle$ in the PDF of $\rm DM_{\rm ext}$. 
	
	For CC data, we apply the Gaussian likelihood 
	\begin{equation} \label{like_CC}
		\log \mathcal{L}(H_i, z_i) \propto -\frac{1}{2} \sum_i \left( \frac{H_{i}^{\rm obs} - H^{\rm th}(z_i)}{\sigma_i} \right)^2\;.
	\end{equation}
	Here, $H_{i}^{\rm obs}$ represents the observational data and $H^{\rm th}$ the cosmological model under consideration, given by Eq. \eqref{H_cosmo} in the realm of cosmography. 
	
	In the case of SNe data, the likelihood function reads as
	\begin{equation} \label{like_SNe}
		\log \mathcal{L}(\mu_i, z_i) \propto -\frac{1}{2} (\mu_{\rm th}(z_i) - \mu_{\rm obs})^T \mathbf{C}^{-1} (\mu_{\rm th}(z_i) - \mu_{\rm obs})\;,
	\end{equation}
	where $\mathbf{C}$ is the covariance matrix, $\mu_{\rm obs}$ is the observed distance modulus from the SNe data, and $\mu_{\rm th}$ is the theoretical distance modulus, given by
	\begin{equation}
		\mu(z) = m_b - M_b = 5 \log_{10} \left( \frac{d_{\rm L}(z)}{1 \; \text{Mpc}} \right) + 25\;,
	\end{equation}
	with $m_b$ and $M_b$ being, respectively, the apparent magnitude and absolute magnitude, and $d_{\rm L}$ is the distance luminosity according to Eq. \eqref{dL_cosmo}.  In cosmological analyses employing type Ia SNe, one typically encounters a strong degeneracy between $M_B$ and $H_0$, because observations of distance modulus constrain only their difference. For the absolute magnitude, we adopt a Gaussian prior with $M_B = -19.2 \pm 0.04~{\rm mag}$, following \citet{camarena2021use}. By imposing such a Gaussian prior on $M_B$, informed by external calibrations (e.g., Cepheid or local distance-ladder measurements), one effectively breaks this degeneracy: the prior anchors $M_B$ to a physically motivated value, thereby translating luminosity-calibration uncertainties into the inference of $H_0$, rather than allowing them to trade off in the posterior arbitrarily. This choice also ensures that local calibration information is properly incorporated and avoids the potential for double-counting low-redshift supernovae, while maintaining a statistically principled Bayesian framework.

	For BAO data, the likelihood function is such that
	\begin{equation} \label{like_BAO}
		\log \mathcal{L}(\mathbf{y}_i, z_i) \propto -\frac{1}{2} (\mathbf{y}_{\rm th}(z_i) - \mathbf{y}_{\rm obs})^T \mathbf{C}^{-1} (\mathbf{y}_{\rm th}(z_i) - \mathbf{y}_{\rm obs})\;,
	\end{equation}
	where $\mathbf{y}_{\rm obs}$ is the set of observed BAO quantities at different redshifts, $\mathbf{C}$ is the covariance matrix provided by the survey, and $\mathbf{y}_{\rm th}(z_i)$ are the corresponding theoretical predictions for the observables $D_M(z)/r_d$, $D_H(z)/r_d$, and $D_V(z)/r_d$, computed either using the equations \eqref{D_M} and \eqref{D_H} all normalized by the sound horizon scale $r_{\rm d}$.  
	
	To break the degeneracy between $H_0$ and $r_{\rm d}$ inherent in uncalibrated BAO measurements, we adopt an external calibration of $r_{\rm d}$ by applying a Gaussian prior of $r_{\rm d} = 147.09 \pm 0.26~{\rm Mpc}$, as inferred from Planck 2018 data under the $\Lambda$CDM framework \citep{aghanim2020planck}. While $r_{\rm d}$ is not a directly measured quantity, its role in the present analysis is to fix the absolute distance scale of BAO observations, rather than to introduce assumptions about the late-time expansion history. The physical processes that set the sound horizon occur in the pre-recombination era and are therefore largely decoupled from the kinematic properties of the low-redshift Universe probed by cosmographic parameters. In this work, the use of a Planck-inferred prior on $r_{\rm d}$ was based on recent late-time, model-independent cosmographic analyses \citep[e.g.,][]{fazzari2025cosmographic}. Nevertheless, we emphasize that this calibration implicitly relies on early-Universe assumptions and does not constitute a fully model-independent determination of the sound horizon. Since cosmographic expansions are only valid at low redshift, they cannot be reliably extrapolated to $z \sim 1100$ in order to include CMB data as an effective BAO point. In this context, calibrating BAO measurements with an externally inferred $r_{\rm d}$ provides a consistent way to anchor our late-time kinematic analysis. Henceforth, we refer to this calibrated BAO combination as DESI+CMB.
	
	Since all events are independent, the combined likelihood of the sample is simply the product of the individual likelihoods: 
	\begin{equation}
		\mathcal{L}_{\rm total} = \prod_{i}\mathcal{L}_i\;\;\;\;\mathrm{or}\;\;\;\;\log \mathcal{L}_{\rm total} = \sum_{i}\log \mathcal{L}_i\;.
	\end{equation}
	So far, we have presented the setup of the likelihoods of individual samples. However, we must combine their likelihoods to include a joint analysis for two or more samples. For instance, for an analysis with FRB, CC, BAO, and SNe data, the joint likelihood is written as 
	\begin{equation}
		\mathcal{L}_{\rm joint} = \mathcal{L}_{\rm FRB} \times \mathcal{L}_{\rm CC} \times \mathcal{L}_{\rm BAO} \times \mathcal{L}_{\rm SNe}\;.
	\end{equation}
	
	Our approach uses Bayesian inference to update our understanding of model parameters based on prior information. This process refers to the dataset $D$ and a probabilistic model for the given distribution $\mathcal{L}(D|\theta)$, known as the likelihood function \citep{John1970,Smith1987}, conditioned on knowledge of the set of adjusted parameters $\theta$. The prior distribution, $\mathcal{P}(\theta)$, quantifies our knowledge of $\theta$. These functions are linked through Bayes' theorem:
	\begin{equation}
		P(\theta|D,\mathcal{M}) = \frac{\mathcal{L}(D|\theta,\mathcal{M}) \cdot \mathcal{P}(\theta|\mathcal{M})}{\mathcal{E}(D|\mathcal{M})}~.
	\end{equation}
	To identify which model best quantifies the data, we need to determine the Bayesian evidence, denoted as $\mathcal{E}(D|\mathcal{M})$, associated with each theoretical model ($\mathcal{M}$) under consideration. To achieve this, we must integrate the likelihood function over the parameter space, as shown below:
	\begin{equation}
		\mathcal{E}(D|\mathcal{M}) = \int \mathcal{L}(D|\theta,\mathcal{M}) \mathcal{P}(\theta|\mathcal{M}) d\theta~.
	\end{equation}
	In this study, we derived posterior probability distributions and the Bayesian evidence with the nested sampling Monte Carlo algorithm MLFriends \citep{buchner2016statistical,buchner2019collaborative} using the \texttt{UltraNest}\footnote{\url{https://johannesbuchner.github.io/UltraNest/}} package \citep{buchner2021ultranest}. Bayesian posterior distributions, along with means, confidence intervals, and contour plots, are computed using the \texttt{GetDist}\footnote{\url{https://getdist.readthedocs.io/en/latest/\#}} package \citep{lewis2019getdist}. This tool employs improved adaptive kernel density estimation techniques, which account for boundary effects and smoothing biases, to generate accurate marginalized densities and visualizations of the parameter space.
	
	\subsection{MCMC analysis steps}
	
	The methodological steps detailed below enable a comprehensive exploration of the cosmological parameter space, allowing for robust constraints across the model. The parameters to be constrained in this analysis are: $\{H_0, q_0, j_0, \Omega_{\rm b}h^2, e^{\mu}, \sigma_{\rm host}, r_d, M_{\rm b}\}$. Table \ref{tab:priors} summarizes the priors used for these parameters.
	
	\begin{table}[ht]
		\centering
		\caption{Priors for the parameters used in the MCMC routine.}
		\begin{tabular}{p{2.5cm} p{1.5cm} p{3.5cm}}
			\hline\hline
			\textbf{Parameter} & \textbf{Prior Type} & \textbf{Range or Mean, Std. Dev.} \\ \hline
			$H_0~[{\rm km/s/Mpc}]$ & Uniform & [60, 80] \\ 
			$r_d~[{\rm Mpc}]$ & Gaussian & $\mu = 147.09$, $\sigma = 0.26$ \\ 
			$e^{\mu}$ & Uniform & [20, 200] \\ 
			$\sigma_{\mathrm{host}}~[\mathrm{pc}~\mathrm{cm}^{-3}]$ & Uniform & [0.2, 2] \\ 
			$M_{\rm b}~[{\rm mag}]$ & Gaussian & $\mu = -19.24$, $\sigma = 0.04$ \\ 
			$\Omega_{\rm b}h^2$ & Uniform & [0.017784, 0.031616] \\ 
			$q_0$ & Uniform & [-0.9, -0.1] \\ 
			$j_0$ & Uniform & [0, 2] \\ \hline\hline
		\end{tabular}
		\label{tab:priors}
	\end{table}
	
	For the cosmography approach, the following steps are performed: 
	
	\begin{itemize}
		\item [(I)] As discussed earlier, for BAO, we adopt an external calibration of $r_{\rm d}$ by applying a Gaussian prior of $r_{\rm d} = 147.09 \pm 0.26~{\rm Mpc}$, as determined by Planck 2018 \citep{aghanim2020planck}. For SNe, we adopt a Gaussian prior with $M_b=-19.2\pm 0.04~{\rm mag}$ for the absolute magnitude according to \citep{camarena2021use}. 
		
		\item [(II)] Flat priors are assumed for $\Omega_{b}h^2$ and $H_0$ with intervals [0.017784, 0.031616] and [60, 80], respectively. The best-fit values of $H_0$, $q_0$, $j_0$, $e^{\mu}$, and $\sigma_{\rm host}$ for FRB, and $H_0$, $q_0$, and $j_0$ for DESI+CMB, SNe and CC are calculated. These parameters are estimated separately for the FRB, SNe, DESI+CMB, and CC datasets.
		
		\item [(III)] Additionally, a joint analysis combining FRB, SNe, DESI+CMB, and CC data is performed to constrain the shared parameters $H_0$, $q_0$, and $j_0$.
	\end{itemize} 
	
	\section{Constraints with cosmography}\label{sec7}
	
\begin{figure}[ht]
		\centering    \includegraphics[width=1.0\linewidth]{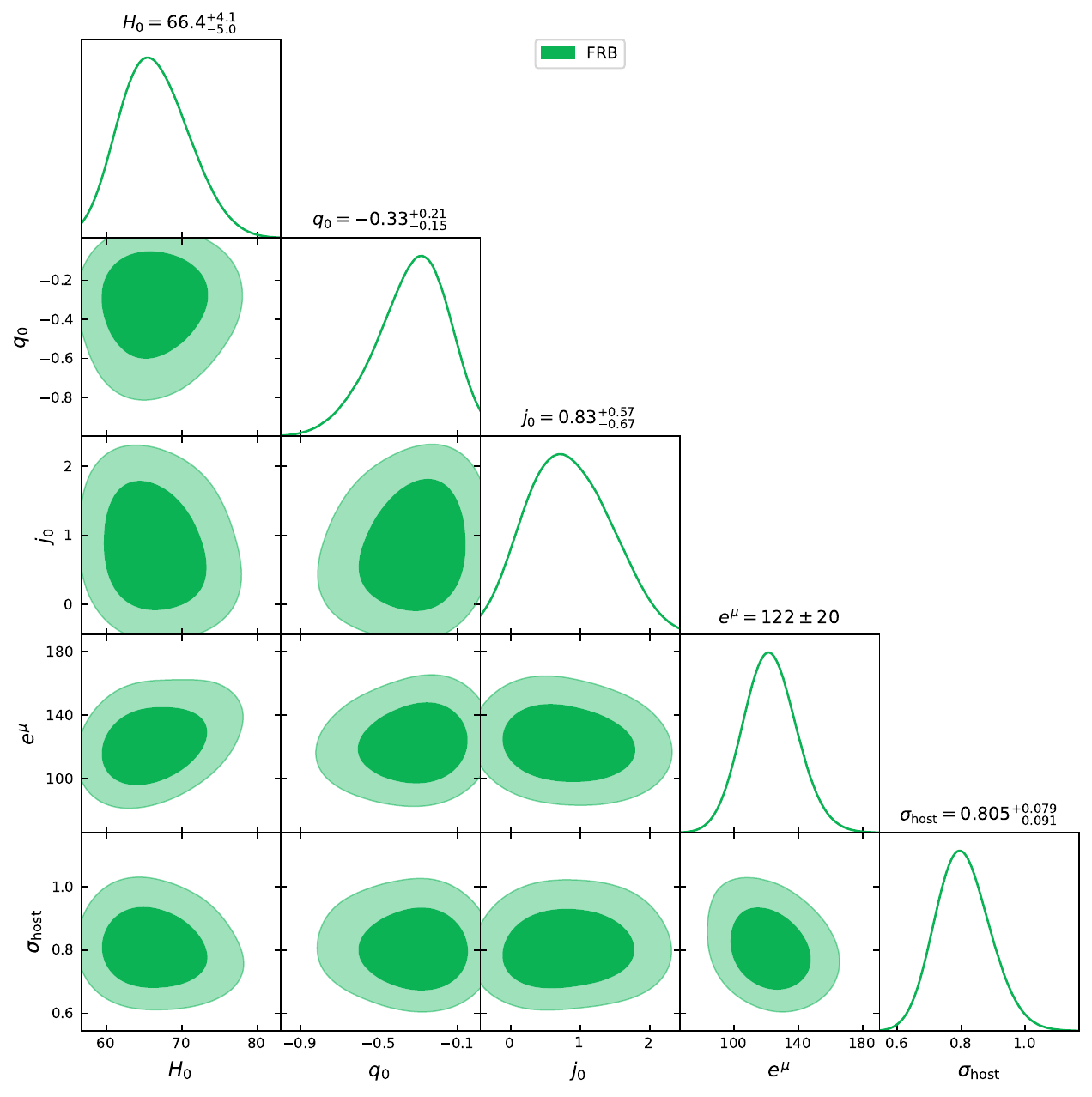}
		\caption{Two-dimensional marginalized contours for the localized FRB sample with $68.3\%$ and $95.4\%$ confidence levels for cosmography. These findings were obtained considering a flat prior for $\Omega_bh^2$ and a fixed value for $f_{\rm IGM}$.}
		\label{fig:4}
	\end{figure}

	In this section, we discuss the findings regarding the cosmographic approach. In the following, we present the results for FRBs alone and then for SNe, DESI+CMB, CC, and their combinations. Fig. \ref{fig:4} unveils the marginalized posterior distributions for the parameters adjusted from the FRB sample. The contours illustrate the one- and two-sigma confidence intervals for the parameters $q_0$, $j_0$, $H_0$, $e^{\mu}$, and $\sigma_{\rm host}$. The marginalized one-dimensional posteriors for each parameter are shown along the diagonal. The first point we should highlight is the Hubble constant obtained through this method, $H_0=66.35_{-5.04}^{+4.13} \mathrm{~km} \, \mathrm{s^{-1}} \, \mathrm{Mpc^{-1}}$, representing a precision of $\sim 6\%$. It is worth noting that the mean is lower than the one from the Planck Collaboration, although the uncertainty lies within the Planck value at a $1\sigma$ confidence level. Notably, this outcome is not compatible with the one found by the SH0ES team. Moreover, the best-fit for the deceleration parameter is $q_0=-0.33_{-0.15}^{+0.21}$, with an accuracy of $\sim 51\%$. For the jerk parameter, we find $j_0=0.83_{-0.67}^{+0.57}$ with $\sim 67\%$ of precision, suggesting that FRBs alone cannot indicate a non-standard cosmology description, favoring the $\Lambda\rm CDM$ model, where $j_0=1.00$. Our results for the cosmographic parameters are compatible within a $1\sigma$ confidence level with those obtained in Ref. \citet{Fortunato:2023deh}.
	
	Additionally, the best-fits for the log-normal distribution parameters are $e^{\mu}=122.21_{-17.10}^{+15.97}$ and $\sigma_{\mathrm{host}}=0.81_{-0.09}^{+0.08}~\mathrm{pc}~\mathrm{cm}^{-3}$, with precisions of $\sim 12\%$, and $\sim 10\%$, respectively. The correlation coefficient between these parameters is $\rho=-0.24$, indicating a weak and statistically insignificant anticorrelation. Moreover, neither $e^{\mu}$ nor $\sigma_{\mathrm{host}}$ shows any appreciable correlation with $H_0$, as seen in Fig. \ref{fig:4}.
	
	\begin{table*}[h]
		\centering
		\caption{Cosmographic parameter constraints ($1\sigma$)}
		\begin{tabular}{l C{2cm} C{1.8cm} C{3cm} C{1.8cm} C{5.6cm}}
			\hline\hline
			\textbf{Parameter} & \textbf{FRB} & \textbf{SNe} & \textbf{DESI+CMB} & \textbf{CC} & \textbf{FRB+SNe+CC+(DESI+CMB)} \\ 
			\hline
			$H_0$      & $66.35_{-5.04}^{+4.13}$ & $73.12_{-1.33}^{+1.35}$ & $65.59_{-1.24}^{+1.25}$ & $68.80_{-3.59}^{+3.66}$ & $68.03_{-0.52}^{+0.53}$ \\
			$q_0$      & $-0.33_{-0.15}^{+0.21}$ & $-0.42_{-0.07}^{+0.08}$ & $-0.29_{-0.08}^{+0.07}$ & $-0.56_{-0.22}^{+0.21}$ & $-0.41\pm 0.02$ \\
			$j_0$      & $0.83_{-0.67}^{+0.57}$  & $0.62_{-0.44}^{+0.36}$  & $0.58_{-0.04}^{+0.03}$  & $1.00_{-0.64}^{+0.65}$  & $0.55\pm 0.02$  \\
			$e^{\mu}$  & $122.21_{-17.10}^{+15.97}$ & -- & -- & -- & -- \\
			$\sigma_{\mathrm{host}}$ & $0.81_{-0.09}^{+0.08}$ & -- & -- & -- & -- \\
			\hline\hline
		\end{tabular}
		\tablefoot{Constraints on cosmographic parameters from individual and joint probes at $68.3\%$ confidence level ($1\sigma$). The best-fit parameters for the log-normal distribution are also shown for FRBs.}
		\label{tab:cosmo_results}
	\end{table*}
	
	To assess the predictive power of the late-time probes considered here in the realm of cosmography, we conducted a joint analysis using data from FRBs, SNe, CC, and DESI+CMB and their combinations. Table \ref{tab:cosmo_results} displays the best-fits for each probe, as well as those for the joint analysis. The individual constraints from SNe, CC, and DESI+CMB unveil distinct levels of precision for the cosmographic parameters. The SNe dataset provides the most precise determination of the Hubble constant, $H_0 = 73.12^{+1.35}_{-1.33} \,\mathrm{~km} \, \mathrm{s^{-1}} \, \mathrm{Mpc^{-1}}$, corresponding to a relative uncertainty of $1.87\%$. DESI+CMB also yields a highly precise estimate, $H_0 = 65.59^{+1.25}_{-1.24}~ \mathrm{~km} \, \mathrm{s^{-1}} \, \mathrm{Mpc^{-1}}$, with a relative uncertainty of $1.89\%$, while the CC data give a less precise result at the $5\%$ level ($H_0 = 68.80_{-3.59}^{+3.66}~ \mathrm{~km} \, \mathrm{s^{-1}} \, \mathrm{Mpc^{-1}}$). For the deceleration parameter, SNe again perform best, achieving $q_0=-0.42_{-0.07}^{+0.08}$ with a relative uncertainty of about $15.94\%$, followed by DESI+CMB with $q_0=-0.29_{-0.08}^{+0.07}$ ($\sim 25\%$) and CC with $q_0=-0.56_{-0.22}^{+0.21}$ ($35\%$). Regarding the jerk parameter, DESI+CMB stands out with an exceptionally tight constraint, $j_0 = 0.58^{+0.03}_{-0.04}$ (relative uncertainty $\sim 6\%$), in contrast to the significantly higher uncertainties from SNe ($j_0 = 0.62^{+0.36}_{-0.44}$ with $60\%$) and CC ($j_0 = 1.00^{+0.65}_{-0.64}$ with $56\%$). Among the considered probes, only the DESI+CMB combination yields a sufficiently precise determination of $j_0$ to suggest a possible tension with the $\Lambda$CDM expectation $j_0 = 1$, within the adopted BAO calibration.
	
	A comparative analysis of the cosmographic parameters $H_0$, $q_0$, and $j_0$ reveals significant improvements in precision when combining all cosmological probes. For the Hubble constant, the joint constraint is $H_0 = 68.03^{+0.53}_{-0.52}\,\mathrm{~km} \, \mathrm{s^{-1}} \, \mathrm{Mpc^{-1}}$, representing a reduction in uncertainty compared to individual probes, achieving a relative uncertainty of $0.76\%$. This result is particularly relevant as it favors the Planck 2018 estimate over the SH0ES determination, reinforcing the consistency with early-Universe measurements. Although small shifts in central values are observed (e.g., $+2.53\%$ vs. FRB, $-1.12\%$ vs. CC, $-6.96\%$ vs. SNe, and $+3.73\%$ vs. DESI+CMB), the joint result consistently provides the tightest constraints. Regarding the deceleration parameter $q_0$, the joint analysis yields $q_0 = -0.41\pm 0.02$ with a relative uncertainty of $6\%$, significantly reduced when compared to FRB ($51\%$), CC ($35\%$), SNe ($15.94\%$), and DESI+CMB ($25\%$). Although the central values show variations (ranging from $-25.31\%$ to $+43.17\%$), these shifts are accompanied by a consistent gain in precision in at least one bound of the credible interval for most probes. For the jerk parameter $j_0$, the joint constraint of $j_0 = 0.55 \pm 0.02$ stands out as the most precise, with a relative uncertainty of just $3.83\%$. This represents a substantial improvement over all individual probes: $67\%$ (FRB), $56\%$ (CC), $60\%$ (SNe), and $6\%$ (DESI+CMB). The joint analysis also reduces both upper and lower uncertainties in most comparisons, even when modest shifts in the central values are present (e.g., $-33\%$ vs. FRB and $-44.47\%$ vs. CC).
	
	A direct comparison between our cosmographic analysis based on DESI DR2 (2025) and the previous results from \citet{luongo2024model} using DESI DR1 (2024) reveals consistent estimates for the cosmographic parameters, with noticeable improvements when combining multiple cosmological probes. The 2024 analysis reported $j_0 = 0.58^{+0.02}_{-0.01}$ for the DESI+CC dataset (with $r_{\rm d} = 144~{\rm Mpc}$), while our updated results using only the DESI DR2 dataset yield $j_0 = 0.58^{+0.03}_{-0.04} $, indicating agreement within uncertainties but with slightly larger error bars, likely due to updated modeling or data systematics. However, our joint analysis incorporating SNe, FRB, CC, and DESI+CMB from DESI DR2 provides significantly tighter constraints, yielding $j_0 = 0.55 \pm 0.02$. Compared to the 2024 DESI+CC results, our updated constraints remain fully consistent within uncertainties, with only a modest shift of about 5\% in the central value of $j_0$.
	
\begin{figure}[h]
		\centering    \includegraphics[width=1.0\linewidth]{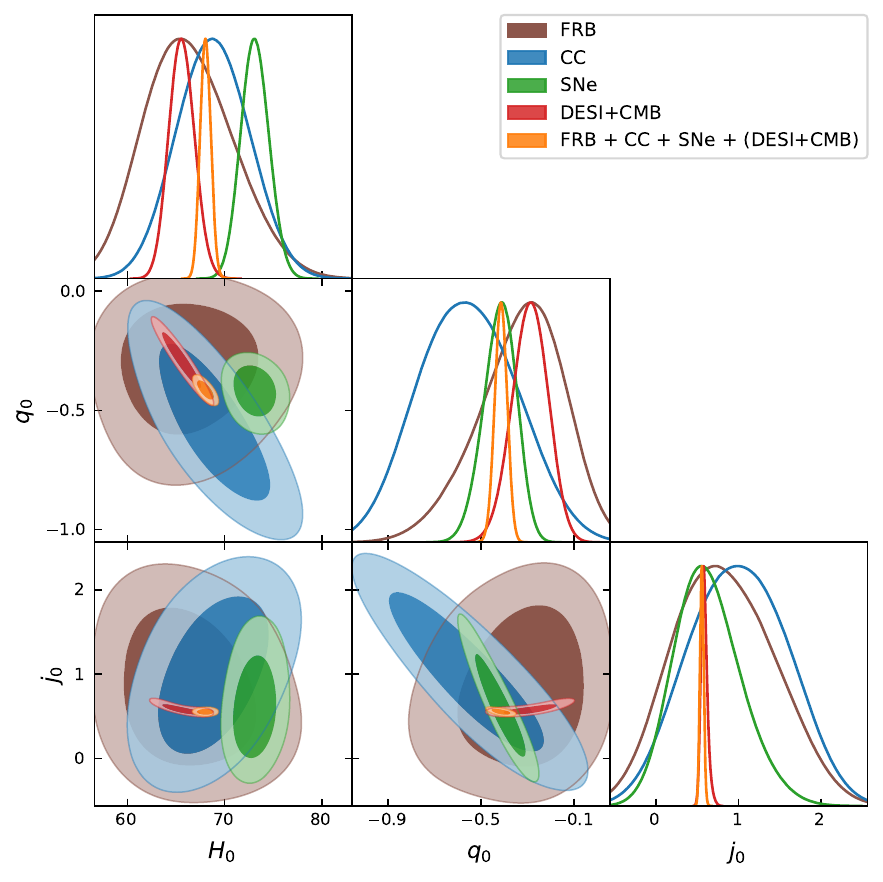}
		\caption{Corner plot exhibiting the marginalized 1D and 2D posterior distributions for the cosmological parameters $H_0$, $q_0$, and $j_0$, and the comparison of several datasets, including the result presented in Fig. \ref{fig:4}, and a joint analysis.}
		\label{fig:5}
	\end{figure}

	A detailed correlation analysis among the cosmographic parameters reveals important insights into the structure and constraints of different cosmological probes. In the FRB-only dataset, we do not find significant correlations between \( H_0 \) and \( q_0 \) (\( \rho = 0.059 \)), \( H_0 \) and \( j_0 \) (\( \rho = -0.15 \)), and between \( q_0 \) and \( j_0 \) (\( \rho = 0.14 \)). The CC dataset exhibits very strong constraints on the cosmographic sector, with a strong negative correlation between \( H_0 \) and \( q_0 \) (\( \rho = -0.76 \)), and a similarly strong anti-correlation between \( q_0 \) and \( j_0 \) (\( \rho = -0.86 \)). A moderate positive correlation is observed between \( H_0 \) and \( j_0 \) (\( \rho = 0.43 \)). The SNe dataset presents a much weaker correlation structure, with \( \rho = -0.21 \) between \( H_0 \) and \( q_0 \), and \( \rho = 0.16\) between \( H_0 \) and \( j_0 \). Nonetheless, the correlation between \( q_0 \) and \( j_0 \) is quite strong and negative (\( \rho = -0.92 \)), reflecting the tight constraint that supernovae impose on the deceleration and jerk parameters. The DESI+CMB probe shows the strongest pairwise correlations among the individual datasets, with \( \rho = -0.95 \) between \( H_0 \) and \( q_0 \), \( \rho = -0.74 \) between \( H_0 \) and \( j_0 \), and \( \rho = 0.67 \) between \( q_0 \) and \( j_0 \). In the combined analysis including all probes (FRB+CC+SNe+(DESI+CMB)), the correlation between \( H_0 \) and \( q_0 \) remains moderately strong and negative (\( \rho = -0.71 \)), while the correlation between \( q_0 \) and \( j_0 \) is still significant (\( \rho = -0.56 \)). The correlation between \( H_0 \) and \( j_0 \) becomes negligible (\( \rho = 0.02 \)), indicating that joint constraints help break degeneracies presented in individual analyses. These results are consistent with the graphical contours shown in Fig. \ref{fig:5}, where DESI+CMB provides the narrowest confidence regions individually, and the joint analysis yields the tightest constraints overall. 
	
	\begin{figure*}[h]
		\centering    
		\includegraphics[width=0.37\linewidth]{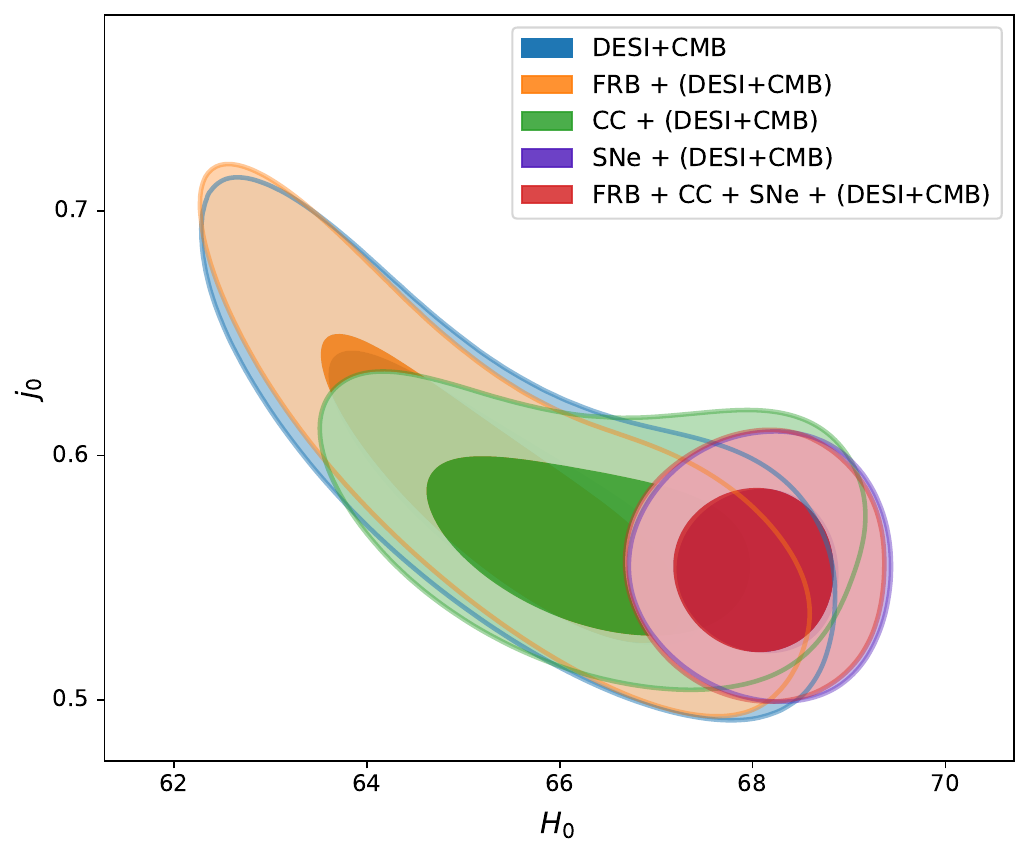} 
		\includegraphics[width=0.37\linewidth]{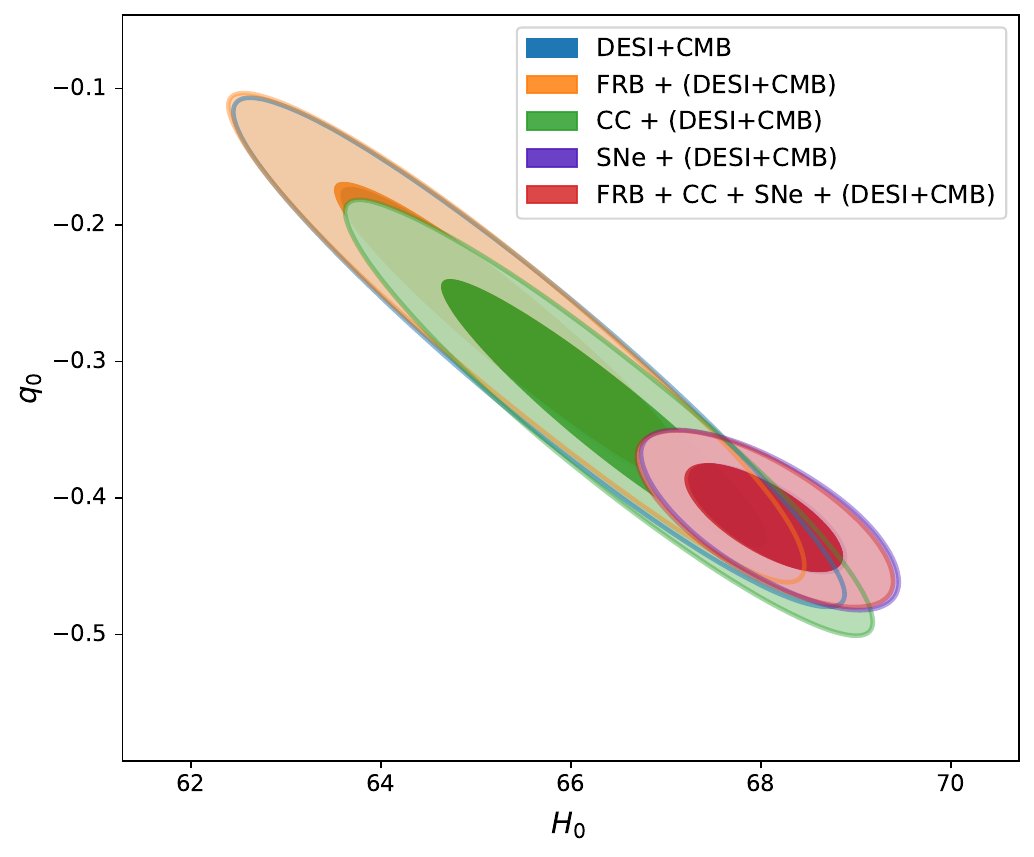}
		\caption{Two-dimensional marginalized contours for the cosmographic parameters. From left to right, the plots depict the confidence regions in the $H_0-j_0$ and $H_0-q_0$ planes, respectively.}
		\label{fig:7}
	\end{figure*}
	
	\begin{figure*}[h]
		\centering    \includegraphics[width=0.37\linewidth]{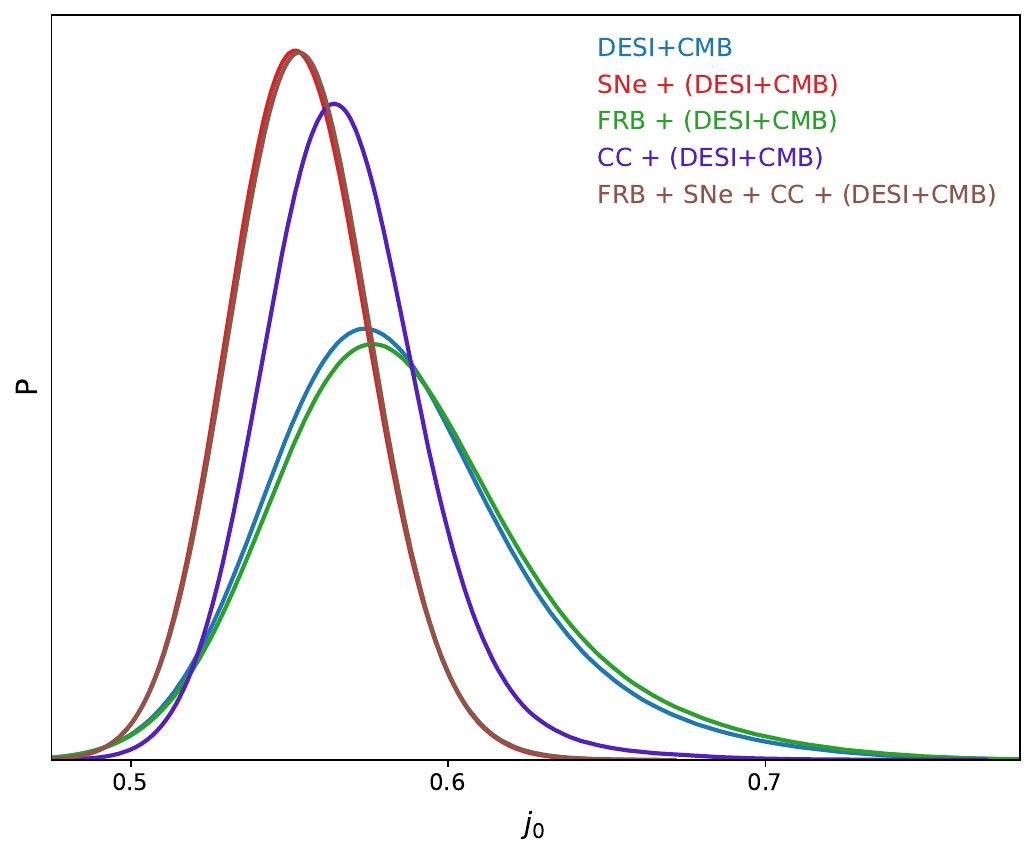}
		\includegraphics[width=0.37\linewidth]{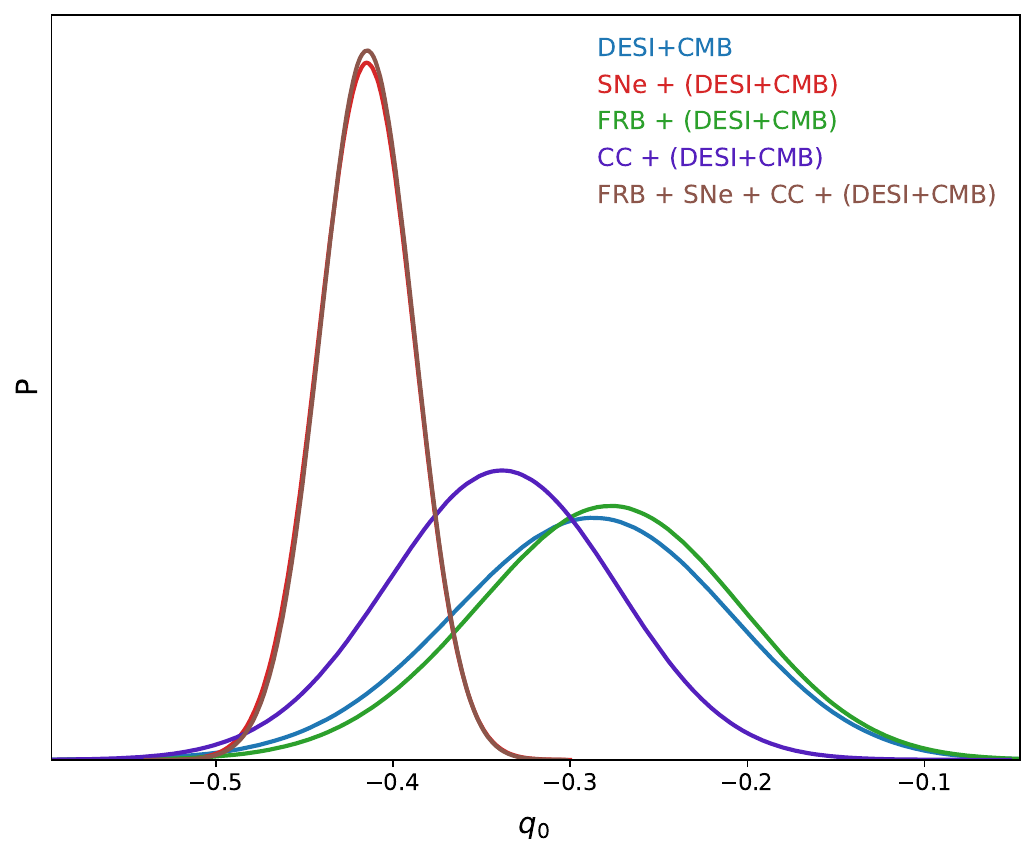}
		\caption{The left (right) panel shows the normalized posterior probability distributions for the jerk (deceleration) parameter.}
		\label{fig:7a}
	\end{figure*}
	
	Fig. \ref{fig:7} portrays the confidence regions showing the correlations or degeneracies between the deceleration and jerk parameters with the Hubble parameter. We evaluated several combinations of DESI+CMB with other datasets (FRB, CC, and SNe) and the joint analysis. The (DESI+CMB)-only contours exhibit tight but elongated degeneracy regions, particularly along the \( H_0 \)–\( j_0 \) direction. When FRB data are included, the contours shrink modestly, indicating that FRBs provide complementary information that helps to alleviate the parameter degeneracy partially. The addition of CC further tightens the constraints, especially along the \( j_0 \) axis, and its impact is substantially stronger than that of FRB. The most noticeable improvement comes from the inclusion of SNe, which significantly reduces the uncertainty in both parameters and modifies the degeneracy direction, resulting in a more circular confidence region.  The full combination (FRB+CC+SNe+(DESI+CMB)) offers only a marginal refinement relative to the SNe+(DESI+CMB) combination, indicating that once SNe are included, additional probes contribute only a slight further improvement. These constraints demonstrate that the synergy between datasets not only reduces uncertainties but also breaks the degeneracies inherent to DESI+CMB alone. In the $H_0-q_0$ plane, the contour lines exhibit a trend similar to that observed in the $H_0-j_0$ plane. However, rather than showing clear degeneracy, they show significant correlations between the parameters. When combining DESI+CMB and FRB data, the correlation between \( H_0 \) and \( q_0 \) remains strong \( \rho = -0.95 \). Similar behaviors are observed in the (DESI+CMB)+CC and (DESI+CMB)+SNe combinations, where DESI+CMB dominates the correlation structure and improves the parameter constraints. 
	
	The left panel of Fig. \ref{fig:7a} displays the one-dimensional normalized posterior probability distributions for the jerk parameter \( j_0 \), comparing DESI+CMB alone with the combined FRB, CC, and SNe datasets. DESI+CMB by itself yields a broad distribution that progressively narrows when additional probes are included. The inclusion of FRB does not lead to improvements in the distribution, suggesting that it is not yet strong enough to determine $j_0$ decisively, as discussed earlier. However, CC causes significant improvements, shifting the distribution slightly to the left and substantially increasing its probability. The inclusion of SNe results in the narrowest and most peaked distribution and greatly tightens the constraints, highlighting its dominance over FRB and CC. The joint analysis does not significantly affect the SNe+(DESI+CMB) combination. Meanwhile, the right panel of Fig. \ref{fig:7a} shows the distributions for the deceleration parameter $q_0$. By taking only the DESI+CMB dataset, one can note a broad distribution that gradually narrows with the inclusion of other probes. Notice that the addition of the FRB dataset slightly shifts the distribution to the right, increasing its probability, indicating a slight improvement in the determination of $q_0$. Similarly to the case of the jerk parameter, adding CC yields a narrower distribution, with a slightly higher probability than FRB. The inclusion of SNe also results in a narrower distribution and greatly tightens the constraints. Lastly, the joint analysis yields the narrowest and most peaked distribution, slightly improving the constraints concerning the SNe+(DESI+CMB) combination.

\begin{figure}[h]
		\centering    \includegraphics[width=1.0\linewidth]{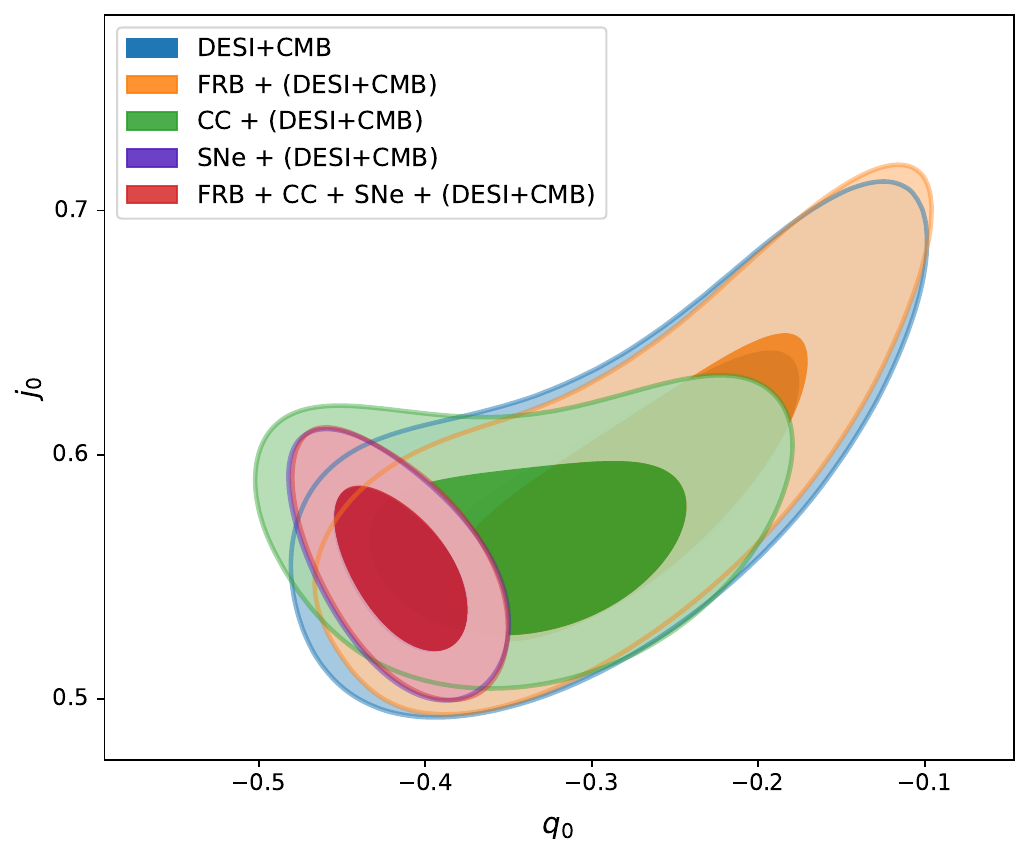}
		\caption{Two-dimensional marginalized contours for the $q_0-j_0$ plane.} 
		\label{fig:8}
	\end{figure}

	Fig. \ref{fig:8} shows the corner plot in the \( q_0 \)-\( j_0 \) plane for DESI+CMB (baseline) and its combinations with additional probes. Similarly to the previous cases, the baseline contours shrink progressively as new datasets are included. In particular, there is a degeneracy between $q_0$ and $j_0$ for DESI+CMB. As can be seen, including FRB in the baseline is not sufficient to alleviate this degeneracy, even if it offers a modest decrease in parameter uncertainty. However, including CC performs marginally better than FRB. Due to the restrictive power of SNe, combining them with the baseline effectively breaks the degeneracy and markedly reduces the uncertainties of the parameters. Once again, the full combination does not significantly affect the dominant SNe+(DESI+CMB) combination.

\begin{figure}[h]
		\centering    \includegraphics[width=1.0\linewidth]{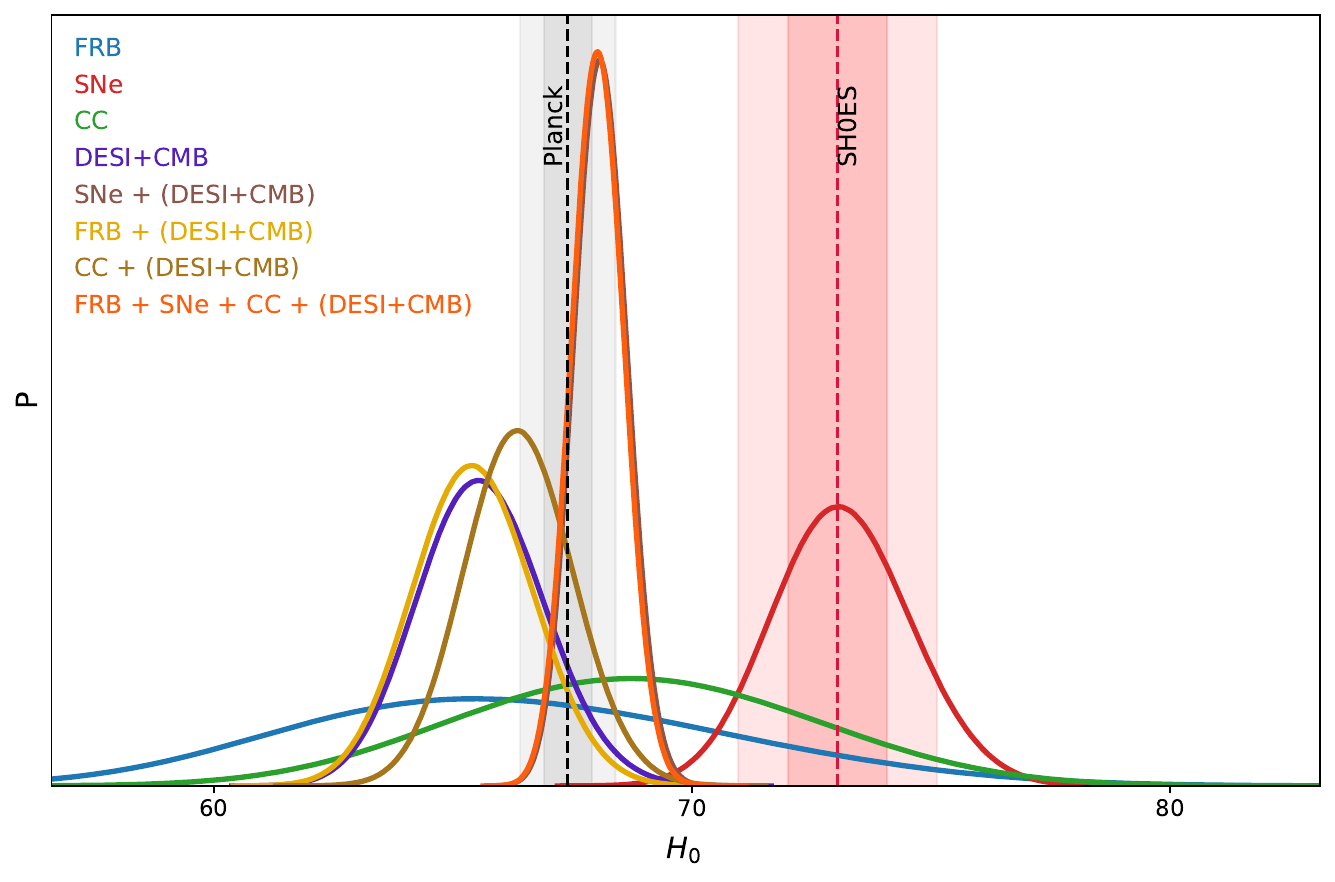}
		\caption{Normalized posterior distributions for the Hubble constant $H_0$ of individual datasets and various combinations of these datasets with DESI+CMB.} 
		\label{fig:6}
	\end{figure}

	A focused comparison of the normalized posterior distributions for the Hubble constant derived from our cosmographic analysis is depicted in Fig. \ref{fig:6}. The (DESI+CMB)-only result exhibits a relatively broad posterior that is not compatible with the Planck measurement at the $1\sigma$ level, and is even more discrepant with the SH0ES estimate. The addition of FRB data induces a mild shift of the distribution toward lower $H_0$ values, slightly increasing its probability, while preserving the overall shape and width of the (DESI+CMB)-alone distribution. The (DESI+CMB)+CC combination leads to a moderately narrower posterior with a higher probability near the Planck value. Notably, the combination with SNe data yields a significant tightening of the constraint, with the peak moving closer to Planck and strongly disfavoring the SH0ES region. The joint analysis further reinforces this trend, producing the most precise estimate, which is in clear agreement with Planck and excludes SH0ES within the $1\sigma$ range. Among the individual probes, FRB and CC present the most significant uncertainties, but their central values also fall closer to the Planck region than to SH0ES. Overall, our cosmographic approach consistently favors lower values of $H_0$, providing independent support to the Planck result and contributing to the broader effort to understand the origin and persistence of the Hubble tension.
	
\begin{figure*}[h]
		\centering    
		\includegraphics[width=1.0\linewidth]{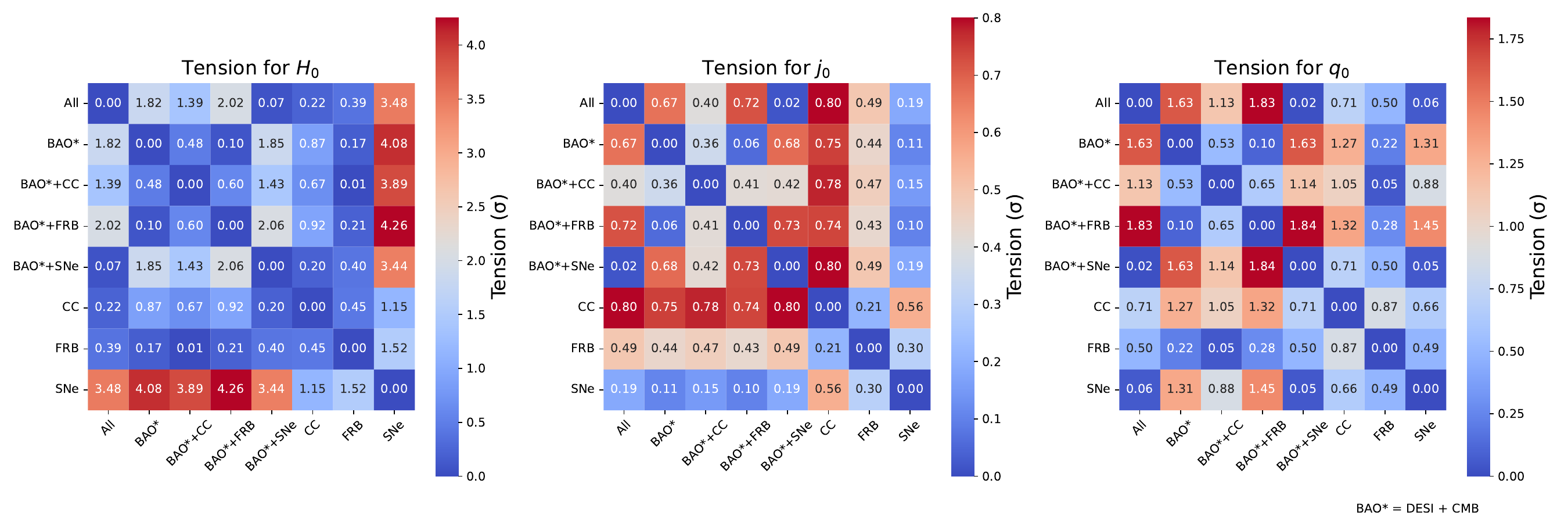}
		\caption{Tension matrices for the cosmographic parameters $H_0$, $j_0$ and $q_0$, comparing the level of disagreement in units of $\sigma$ between different dataset combinations. The label $\mathrm{BAO}^*$ represents the combination DESI+CMB.}
		\label{fig:9}
	\end{figure*}

	To quantify the tension between two independent measurements of a cosmological parameter, we use the standard tension metric defined as (in units of standard deviations $\sigma$):
	\begin{equation}
		T = \frac{|\mu_1 - \mu_2|}{\sqrt{\sigma_1^2 + \sigma_2^2}}~,
	\end{equation}
	where \( \mu_1 \) and \( \mu_2 \) are the means, and \( \sigma_1 \) and \( \sigma_2 \) are the standard deviations of the two measurements, respectively. The values of \( T \gtrsim 3\sigma \) are typically interpreted as indicating significant tension. Fig. \ref{fig:9} presents the tension matrices to evaluate the discrepancy between the cosmographic parameters. In the left panel, the Hubble constant ($H_0$) exhibits a significant tension between SNe and DESI+CMB, with a discrepancy of $\sim 4.1\sigma$ in the tension matrix. This tension is compatible with that found by the DESI DR2 collaboration ($\sim 4.5\sigma$). The middle panel shows the jerk parameter ($j_0$) with tensions between datasets, such as SNe vs. FRB ($\sim 0.3\sigma$) and DESI+CMB vs. FRB ($\sim 0.4\sigma$). In the right panel, the deceleration parameter ($q_0$) displays moderate tensions (e.g., $1.31\sigma$ for SNe vs. DESI+CMB), lower than those for $H_0$, suggesting better agreement across datasets.
	
	\section{Conclusions}
	\label{conclusion}
	
	Late-time observational probes provide a powerful and complementary framework for constraining the kinematics of cosmic expansion. In this work, we have combined multiple late-time probes, including fast radio bursts, within a cosmographic approach valid at low redshift ($z \lesssim 1$). This method employs a Taylor series expansion of the scale factor to derive kinematic parameters ($H_0$, $q_0$, $j_0$) directly from observational data, thereby bypassing assumptions about the universe's composition. It can mitigate model dependencies by leveraging datasets that include 114 localized FRBs, 1563 SNe Ia light curves, 7 DESI DR2 BAO measurements, and 27 CC data points to achieve robust constraints. The integration of these probes, combined through Bayesian inference with likelihood functions tailored to each dataset, enhances the precision of our results and addresses degeneracies inherent in individual probes.
	
	Our findings demonstrate that FRBs alone can constrain the Hubble constant with precision of $\sim 6\%$. The cosmographic approach estimates $H_0=66.35_{-5.04}^{+4.13} \mathrm{~km} \, \mathrm{s^{-1}} \, \mathrm{Mpc^{-1}}$ which is incompatible with the SH0ES forecast at a 1$\sigma$ confidence level, being closer to the early-Universe inference from Planck. Besides, the FRB-only analysis also yields $q_0 = -0.33^{+0.21}_{-0.15}$ and $j_0 = 0.83^{+0.57}_{-0.67}$, both fully consistent with the $\Lambda$CDM expectation within $1\sigma$. Notably, the DESI+CMB probe provides a precise determination of the Hubble constant, $H_0 = 65.59^{+1.25}_{-1.24}~\mathrm{km\,s^{-1}\,Mpc^{-1}}$, at the $1.89\%$ level. The deceleration and jerk parameters found were $q_0 = -0.29^{+0.07}_{-0.08}$ and $j_0 = 0.58^{+0.03}_{-0.04}$. Within the context of the adopted sound horizon calibration, the value of $j_0$ departs from the $\Lambda$CDM expectation ($j_0 = 1$) at the $1\sigma$ level, suggesting a possible tension in the late-time kinematic sector. A joint analysis combining FRB, SNe, DESI+CMB, and CC datasets yields substantially tighter constraints: 
	$H_0 = 68.03^{+0.53}_{-0.52}\,\mathrm{km\,s^{-1}\,Mpc^{-1}}$, corresponding to a relative uncertainty of $0.76\%$, 
	$q_0 = -0.41 \pm 0.02$, and 
	$j_0 = 0.55 \pm 0.02$. In this last scenario, the inferred value of the jerk parameter remains lower than the $\Lambda$CDM value. The DESI+CMB dataset primarily drives this trend.
	
	Looking ahead, the ongoing increase in the number and precision of well-localized FRBs, supported by upcoming surveys and new instruments, will be crucial to improving the statistical significance of our constraints. Although current limitations in sample size still pose challenges, leading to comparatively low precision in the determination of $H_0$ relative to other late-time and early-Universe estimates, the demonstrated complementarity between FRBs and other probes opens a promising path for precision cosmology. As larger and more accurate FRB datasets become available, both from existing radio telescopes and the new generation, such as the BINGO radio telescope \citep{dosSantos:2023ipw}, we expect meaningful improvements in our ability to constrain not only the Hubble constant but also higher-order cosmographic parameters. This progress will enable increasingly stringent reconstructions of the late-time expansion history and more robust consistency tests of the $\Lambda$CDM kinematic expectations.

\begin{acknowledgements}
      The authors thank Prof. Adam Riess for addressing an important issue regarding the use of DESI data in the determination of $H_0$ values in the first version of this work and an anonymous referee for valuable comments. A.R.Q. work is supported by FAPESQ-PB. A.R.Q. acknowledges support by CNPq under process number 310533/2022-8. L.L.S., K.E.L.F., and R.A.B. thank the Paraíba State Research Support Foundation (FAPESQ) for financial support. J.R.L.S acknowledges support from CNPq (Grants 309494/2021-4 and 302190/2025-2), FAPESQ-PB (Grant 1356/2024), CAPES Finance Code 001 and Alexander von Humboldt-Stiftung Foundation. A. R. M. O. and J. V. thank CAPES. L. F. S. work is supported by CNPq. C.A. W. thanks CNPq for grants 407446/2021-4 and 312505/2022-1, the Brazilian Ministry of Science, Technology and Innovation (MCTI) and the Brazilian Space Agency (AEB), which supported the present work under the PO 20VB.0009.
\end{acknowledgements}

\bibliographystyle{aa} 
\bibliography{bibli}

\begin{@twocolumnfalse}
\onecolumn

\begin{appendix}
\section{Fast radio bursts dataset}\label{append_a}

\begin{longtable}{lccc>{\raggedright\arraybackslash}p{5cm}}

\caption{Fast radio bursts dataset.} \label{tab1} \\
\hline\hline
FRB (name) & Redshift $z$ & DM$_{\rm obs}$ (pc cm$^{-3}$) & DM$_{\rm MW}$ (pc cm$^{-3}$) & Reference \\
\hline
\endfirsthead
\caption{continued}\\
\hline\hline
FRB (name) & Redshift $z$ & DM$_{\rm obs}$ (pc cm$^{-3}$) & DM$_{\rm MW}$ (pc cm$^{-3}$) & Reference \\
\hline
\endhead
\hline \multicolumn{5}{r}{{Continued on next page}} \\
\endfoot
\hline
\endlastfoot
20200120E & 0.0008 & 87.77 & 30 & \cite{Bhardwaj:2021xaa}\\
20181030A & 0.0039 & 103.5 & 40 & \cite{Bhardwaj:2021hgc,CHIMEFRB:2021srp}\\
20171020A & 0.00867 & 114.1 & 38 & \cite{Mahony:2018ddp}\\
20231229A & 0.0190 & 198.5 & 58.12 & \cite{CHIMEFRB:2025ggb}\\
20240210A & 0.0238 & 283.75 & 31 & \cite{Shannon:2024pbu}\\
20181220A & 0.027 & 209.4 & 125.8 & \cite{Bhardwaj:2023vha,CHIMEFRB:2021srp}\\
20231230A & 0.0298 & 131.4 & 61.51 & \cite{CHIMEFRB:2025ggb}\\
20181223C & 0.03024 & 112.51 & 19.91 & \cite{Bhardwaj:2023vha}\\
20190425A & 0.03122 & 128.16 & 48.75 & \cite{Bhardwaj:2023vha,CHIMEFRB:2021srp}\\
20180916B & 0.0337 & 348.76 & 200 & \cite{Gordon:2023cgw,CHIMEFRB:2021srp}\\
20230718A & 0.035 & 477 & 396 & \cite{glowacki2024h}\\
20240201A & 0.0427 & 374.5 & 38 & \cite{Shannon:2024pbu}\\
20220207C & 0.0430 & 262.38 & 79.3 & \cite{Law:2023ibd}\\
20211127I & 0.0469 & 234.83 & 42.5 & \cite{Gordon:2023cgw}\\
20201123A & 0.0507 & 433.55 & 251.93 & \cite{Rajwade:2022zkj}\\
20230926A & 0.0553 & 222.8 & 52.69 & \cite{CHIMEFRB:2025ggb}\\
20200223B & 0.06024 & 202.268 & 46 & \cite{Ibik:2023ugl,CHIMEFRB:2021srp}\\
20190303A & 0.064 & 222.4 & 26 & \cite{Michilli:2022bbs,CHIMEFRB:2021srp}\\
20231204A & 0.0644 & 221.0 & 29.73 & \cite{CHIMEFRB:2025ggb}\\
20231206A & 0.0659 & 457.7 & 59.13 & \cite{CHIMEFRB:2025ggb}\\
20210405I & 0.066 & 565.17 & 516.1 & \cite{Driessen:2023lxj}\\
20180814 & 0.068 & 189.4 & 87 & \cite{Michilli:2022bbs,CHIMEFRB:2021srp}\\
20231120A & 0.07 & 438.9 & 43.8 & \cite{DeepSynopticArrayTeam:2023iev,Sharma:2024fsq,Connor:2024mjg}\\
20231005A & 0.0713 & 189.4 & 33.37 & \cite{CHIMEFRB:2025ggb}\\
20190418A & 0.07132 & 184.5 & 70.1 & \cite{Bhardwaj:2023vha,CHIMEFRB:2021srp}\\
20211212A & 0.0715 & 206.0 & 27.1 & \cite{Gordon:2023cgw}\\
20231123A & 0.0729 & 302.1 & 89.76 & \cite{CHIMEFRB:2025ggb}\\
20220912A & 0.0771 & 219.46 & 115 & \cite{DeepSynopticArrayTeam:2022rbq,Zhang:2023eui}\\
20231011A & 0.0783 & 186.3 & 70.36 & \cite{CHIMEFRB:2025ggb}\\
20220509G & 0.0894 & 269.53 & 55.2 & \cite{Law:2023ibd}\\
20230124 & 0.0940 & 590.6 & 38.5 & \cite{DeepSynopticArrayTeam:2023iev,Sharma:2024fsq,Connor:2024mjg}\\
20201124A & 0.098 & 413 & 123 & \cite{Gordon:2023cgw,Lanman:2021yba}\\
20230708A & 0.105 & 411.51 & 50 & \cite{Shannon:2024pbu}\\
20231223C & 0.1059 & 165.8 & 47.9 & \cite{CHIMEFRB:2025ggb}\\
20191106C & 0.10775 & 333.4 & 25 & \cite{Ibik:2023ugl,CHIMEFRB:2021srp}\\
20231128A & 0.1079 & 331.6 & 25.05 & \cite{CHIMEFRB:2025ggb}\\
20230222B & 0.11 & 187.8 & 27.7 & \cite{CHIMEFRB:2025ggb}\\
20231201A & 0.1119 & 169.4 & 70.03 & \cite{CHIMEFRB:2025ggb}\\
20220914A & 0.1139 & 631.28 & 55.2 & \cite{Law:2023ibd}\\
20190608B & 0.1178 & 339 & 37 & \cite{Gordon:2023cgw,Hiramatsu:2022tyn}\\
20230703A & 0.1184 & 291.3 & 26.97 & \cite{CHIMEFRB:2025ggb}\\
20240213A & 0.1185 & 357.4 & 40.1 & \cite{Connor:2024mjg}\\
202030222A & 0.1223 & 706.1 & 134.13 & \cite{CHIMEFRB:2025ggb}\\
20190110C & 0.1224 & 221.961 & 35.66 & \cite{Ibik:2023ugl,CHIMEFRB:2021srp}\\
20230628A & 0.1265 & 345.15 & 39.1 & \cite{DeepSynopticArrayTeam:2023iev,Sharma:2024fsq,Connor:2024mjg}\\
20240310A & 0.127 & 601.8 & 36 & \cite{Shannon:2024pbu}\\
20210807D & 0.1293 & 251.9 & 121.2 & \cite{Gordon:2023cgw}\\
20240114A & 0.13 & 527.65 & 49.7 & \cite{Kumar:2024svu}\\
20240209A & 0.1384 & 176.49 & 55.5 & \cite{Shah:2024ywp}\\
20210410D & 0.1415 & 578.78 & 56.2 & \cite{Gordon:2023cgw,Caleb:2023atr}\\
20230203A & 0.1464 & 420.1 & 36.29 & \cite{CHIMEFRB:2025ggb}\\
20231226A & 0.1569 & 329.9 & 145 & \cite{Shannon:2024pbu}\\
20230526A & 0.157 & 316.4 & 50 & \cite{Shannon:2024pbu}\\
20220920A & 0.158 & 314.99 & 40.3 & \cite{Law:2023ibd}\\
20200430A & 0.1608 & 380.25 & 27 & \cite{Gordon:2023cgw,Hiramatsu:2022tyn}\\
20210603A & 0.177 & 500.15 & 40 & \cite{Cassanelli:2023hvg}\\
20220529A & 0.1839 & 246.0 & 40.0 & \cite{gao2024measuring}\\
20230311A & 0.1918 & 364.3 & 92.39 & \cite{CHIMEFRB:2025ggb}\\
20220725A & 0.1926 & 290.4 & 31 & \cite{Shannon:2024pbu}\\
20121102A & 0.19273 & 557.0 & 188.4 & \cite{Gordon:2023cgw}\\
20221106A & 0.2044 & 343.8 & 35 & \cite{Shannon:2024pbu}\\
20240215A & 0.21 & 549.5 & 48.0 & \cite{Connor:2024mjg}\\
20230730A & 0.2115 & 312.5 & 85.18 & \cite{CHIMEFRB:2025ggb}\\
20210117A & 0.214 & 729.0 & 34.0 & \cite{Bhandari:2022ton}\\
20221027A & 0.229 & 452.5 & 47.2 & \cite{Connor:2024mjg}\\
20191001A & 0.234 & 506.92 & 44.7 & \cite{Gordon:2023cgw,Bhandari:2020cde}\\
20190714A & 0.2365 & 504.13 & 38 & \cite{Gordon:2023cgw,Hiramatsu:2022tyn,HESS:2021smp,Guidorzi:2020ggq}\\
20221101B & 0.2395 & 490.7 & 131.2 & \cite{DeepSynopticArrayTeam:2023iev,Sharma:2024fsq,Connor:2024mjg}\\
20220825A & 0.2414 & 651.24 & 79.7 & \cite{Law:2023ibd}\\
20190520B & 0.2418 & 1204.7 & 60.2 & \cite{Gordon:2023cgw}\\
20191228A & 0.2432 & 297.5 & 33 & \cite{Bhandari:2021pvj}\\
20231017A & 0.2450 & 344.2 & 64.55 & \cite{CHIMEFRB:2025ggb}\\
20221113A & 0.2505 & 411.4 & 91.7 & \cite{DeepSynopticArrayTeam:2023iev,Sharma:2024fsq,Connor:2024mjg}\\
20220307B & 0.2507 & 499.15 & 128.2 & \cite{Law:2023ibd}\\
20220831A & 0.262 & 1146.25 & 126.7 & \cite{Connor:2024mjg}\\
20231123B & 0.2625 & 396.7 & 40.2 & \cite{DeepSynopticArrayTeam:2023iev,Sharma:2024fsq,Connor:2024mjg}\\
20230307A & 0.2710 & 608.9 & 37.6 & \cite{DeepSynopticArrayTeam:2023iev,Sharma:2024fsq,Connor:2024mjg}\\
20221116A & 0.2764 & 640.6 & 132.3 & \cite{Sharma:2024fsq,Connor:2024mjg}\\
20220105A & 0.2785 & 583 & 22 & \cite{Gordon:2023cgw}\\
20210320C & 0.2796 & 384.8 & 42.2 & \cite{Gordon:2023cgw}\\
20221012A & 0.2846 & 441.08 & 54.4 & \cite{Law:2023ibd}\\
20240229A & 0.287 & 491.15 & 37.9 & \cite{Connor:2024mjg}\\
20190102C & 0.2913 & 363.6 & 57.3 & \cite{Gordon:2023cgw}\\
20220506D & 0.3004 & 396.97 & 89.1 & \cite{Law:2023ibd}\\
20230501A & 0.3010 & 532.5 & 125.6 & \cite{DeepSynopticArrayTeam:2023iev,Connor:2024mjg}\\
20180924B & 0.3214 & 361.42 & 40.5 & \cite{Gordon:2023cgw}\\
20231025B & 0.3238 & 368.7 & 48.67 & \cite{CHIMEFRB:2025ggb}\\
20230626A & 0.3270 & 451.2 & 39.2 & \cite{DeepSynopticArrayTeam:2023iev,Sharma:2024fsq,Connor:2024mjg}\\
20180301A & 0.3304 & 536 & 152 & \cite{Gordon:2023cgw,Price:2019fmc}\\
20231220A & 0.3355 & 491.2 & 49.9 & \cite{Connor:2024mjg}\\
20211203C & 0.3439 & 635.0 & 63.4 & \cite{Gordon:2023cgw}\\
20220208A & 0.3510 & 437.0 & 101.6 & \cite{Sharma:2024fsq,Connor:2024mjg}\\
20220726A & 0.3610 & 686.55 & 89.5 & \cite{DeepSynopticArrayTeam:2023iev,Sharma:2024fsq,Connor:2024mjg}\\
20220717A & 0.36295 & 637.34 & 118.3 & \cite{Rajwade:2024ozu}\\
20230902A & 0.3619 & 440.1 & 34 & \cite{Shannon:2024pbu}\\
20200906A & 0.3688 & 577.8 & 36 & \cite{Gordon:2023cgw,Hiramatsu:2022tyn}\\
20240119A & 0.37 & 483.1 & 37.9 & \cite{Connor:2024mjg}\\
20220330D & 0.3714 & 468.1 & 38.6 & \cite{Sharma:2024fsq,Connor:2024mjg}\\
20190611B & 0.3778 & 321.4 & 57.8 & \cite{Gordon:2023cgw}\\
20220501C & 0.381 & 449.5 & 31 & \cite{Shannon:2024pbu}\\
20220204A & 0.4 & 612.2 & 50.7 & \cite{DeepSynopticArrayTeam:2023iev,Sharma:2024fsq,Connor:2024mjg} \\
20230712A & 0.4525 & 586.96 & 39.2 & \cite{DeepSynopticArrayTeam:2023iev,Sharma:2024fsq,Connor:2024mjg}\\
20181112A & 0.4755 & 589.27 & 42 & \cite{Gordon:2023cgw}\\
20220310F & 0.4779 & 462.24 & 45.4 & \cite{Law:2023ibd} \\
20220918A & 0.491 & 656.8 & 41 & \cite{Shannon:2024pbu}\\
20190711A & 0.5220 & 593.1 & 56.4 & \cite{Gordon:2023cgw,Macquart:2020lln}\\
20230216A & 0.5310 & 828.0 & 38.5 & \cite{Sharma:2024fsq,Connor:2024mjg}\\
20230814A & 0.5535 & 696.4 & 104.9 & \cite{Connor:2024mjg} \\
20221219A & 0.5540 & 706.7 & 44.4 & \cite{DeepSynopticArrayTeam:2023iev,Sharma:2024fsq,Connor:2024mjg}\\
20190614D & 0.60 & 959.2 & 83.5 & \cite{Law:2020cnm,Hiramatsu:2022tyn}\\
20220418A & 0.6220 & 623.25 & 37.6 & \cite{Law:2023ibd}\\
20190523A & 0.6600 & 760.8 & 37 & \cite{Ravi:2019alc}\\
20240123A & 0.968 & 1462.0 & 90.3 & \cite{Connor:2024mjg}\\
20221029A & 0.9750 & 1391.05 & 43.9 & \cite{DeepSynopticArrayTeam:2023iev,Sharma:2024fsq,Connor:2024mjg} \\
20220610A & 1.016 & 1458.1 & 30.9 & \cite{Ryder:2022qpg}\\
20230521B & 1.354 & 1342.9 & 138.8 & \cite{Shannon:2024pbu,Connor:2024mjg}\\
\end{longtable}

\end{appendix}

\twocolumn
\end{@twocolumnfalse}

\end{document}